\documentclass[twocolumn,english,prl,showpacs,superscriptaddress,longbibliography]{revtex4-1}
\usepackage{amsfonts}
\usepackage{amsmath}
\usepackage{amssymb}
\usepackage[colorlinks, citecolor=blue,linkcolor=red]{hyperref}
\usepackage{color}
\usepackage{hyperref}
\usepackage{textcomp}
\usepackage[rightcaption]{sidecap}
\usepackage{subfigure}
\usepackage[rightcaption]{sidecap}
\usepackage[utf8]{inputenc}
\usepackage[T1]{fontenc}
\usepackage{graphicx}
\usepackage{lipsum}

\newcommand{\st}{\mathrm{st}}
\newcommand{\eff}{\mathrm{eff}}
\newcommand{\gb}{\mathrm{gb}}
\newcommand{\ret}{\mathrm{ret}}
\newcommand{\dd}{\mathrm{d}}

\begin{document}
	\title{Colored-Noise-Induced Horizon Fluctuations Near a Double Root in an Effective Black-Hole Geometry}
	\author{Kashif Ammar Yasir}
	\email{kayasir@zjnu.edu.cn}
	\affiliation{Department of Physics, Zhejiang Normal University, Jinhua 321004, China.}
	\affiliation{Zhejiang Institute of Photoelectronics, Jinhua 321004, China.}
	\setlength{\parskip}{0pt}
	\setlength{\belowcaptionskip}{-10pt}
	
\begin{abstract}
	Stress-tensor fluctuations carry information that is absent from the mean
	semiclassical Einstein equation, but their observable effect depends as much
	on the response of the geometry as on the noise itself.  We formulate an
	effective, quasistationary spherical model organized by the response--noise
	structure of a quantum Langevin equation.  A prescribed mean dressing and a
	horizon-localized colored source are varied independently.  For every source
	realization we reconstruct the lapse, locate its outer trapping horizon, and
	evaluate the adiabatic Hayward--Kodama temperature from the slope at that same
	root.  As the mean lapse approaches an inner--outer root merger, its inverse
	slope acts as a geometric susceptibility: a fixed source covariance produces
	enhanced horizon fluctuations, horizon--temperature covariance, and a
	positively skewed, greybody-filtered luminosity.  The specific advance is a
	single-realization map from colored metric noise to correlated geometric,
	thermal, and radiative statistics, together with an explicit separation of
	reservoir strength from near-critical amplification.  The model is a
	stochastic-semiclassical testbed, not a microscopic evaluation of Unruh-state
	response and noise kernels.
\end{abstract}

\maketitle

Black-hole thermodynamics is usually stated in terms of mean quantities.
The semiclassical Einstein equation is sourced by the renormalized
expectation value of the stress tensor; the horizon, surface gravity,
temperature, and mean flux are then read from the resulting geometry
\cite{Bardeen1973,Hawking1975,Page1976,Unruh1976,BirrellDavies1982}.  A quantum
state contains more information than this one-point function.  Its connected
stress-tensor correlator drives induced metric fluctuations
\cite{MartinVerdaguer1999,PhillipsHu2001,HuVerdaguer2008}, and earlier
stochastic-gravity work showed that such fluctuations complicate the very
notion of a sharply localized evaporating horizon \cite{HuRoura2007}.  What a
mean-field treatment cannot determine are the joint statistics of observables
that probe different local features of the same random geometry.  The
position of a horizon depends on a zero of the lapse, whereas its adiabatic
temperature depends on the derivative at that zero.  Their widths and
covariance, and the higher moments produced by mapping them into radiation,
therefore require a realization-level calculation.

Open quantum systems provide a useful organization of this problem.  After
environmental modes are eliminated, quantum Brownian motion, input--output
theory, and cavity optomechanics lead to a Langevin equation such as
\(M\ddot Q+V'(Q)+\int^t\dd s\,\gamma(t-s)\dot Q(s)=\hat F(t)\)
\cite{FeynmanVernon1963,CaldeiraLeggett1983,BreuerPetruccione2002,
	Weiss2012,GardinerCollett1985,Clerk2010,Aspelmeyer2014}.  The response kernel
and fluctuating force originate in the same coupling.  Stochastic
semiclassical gravity has an analogous structure: the metric is the system,
the quantum field is the environment, and a closed-time-path influence action
generates a retarded stress-polarization kernel and a stress-tensor noise
kernel \cite{SinhaRavalHu2003,HuRoura2007}.  For an evaporating black hole,
however, the Unruh state is not a global equilibrium bath.  Its mean stress,
causal response, and point-separated noise kernel must be calculated from the
field theory rather than inferred from an equilibrium
fluctuation--dissipation formula \cite{Eftekharzadeh2012}.  The open-system
viewpoint fixes the architecture of the reduced description, as it does in
recent work on stochastic inflation \cite{Li2026}; it does not supply those
black-hole kernels.

In this Letter we use that architecture to build an effective,
quasistationary stochastic geometry.  A power-law exterior density labels a
family of mean lapses, while a horizon-localized colored Gaussian source
represents a projected environmental fluctuation.  Varying the mean dressing
\(\epsilon\) independently of the stochastic multiplier \(\lambda_s\)
separates the susceptibility of the geometry from the imposed source
amplitude.  Each realization is propagated through the same sequence:
radial mass constraint, stochastic lapse, outer trapping-horizon root,
Hayward--Kodama temperature, and greybody-filtered luminosity.  The
\(\epsilon=1100\) branch lies close to an inner--outer root merger.  There the
small mean lapse slope amplifies a fixed colored perturbation and produces
strong horizon--temperature covariance and radiative skewness.  The new
element is not horizon broadening by itself, but the explicit separation of
source covariance from root susceptibility and the resulting joint
geometric--thermal--radiative statistics.  No microscopic Unruh-state kernel
is assumed.

\emph{Open-system reduction.---}
At the level needed for this construction, the linear parent equation is
\begin{equation}
	\begin{split}
		{\cal L}_{\mu\nu}{}^{\alpha\beta}h_{\alpha\beta}^{(\xi)}(x)
		&+\frac{1}{2}\int\dd^4x'\,
		H_{\mu\nu}{}^{\alpha\beta,\ret}(x,x')\\
		&\hspace{0.8cm}\times h_{\alpha\beta}^{(\xi)}(x')
		=8\pi G\,\Xi_{\mu\nu}(x).
	\end{split}
	\label{eq:EL-main}
\end{equation}
Here \({\cal L}\) is the local linearized gravitational operator and
\(H^{\ret}\) is the causal environmental response.  The stochastic source
has zero mean and covariance
\(\langle\Xi_{\mu\nu}(x)\Xi_{\alpha\beta}(x')\rangle_{\st}
=N_{\mu\nu\alpha\beta}(x,x')\).  Formally, the induced metric covariance is
\(\langle hh\rangle_{\st}=(8\pi G)^2G^{\ret}\!\star N\star G^{\rm adv}\),
the gravitational analogue of the susceptibility--noise--susceptibility
relation.  The Supplemental Material derives this connection from both the
oscillator-bath quantum Langevin equation and the closed-time-path influence
functional \cite{Supplemental}.  Equation~\eqref{eq:EL-main} is the formal
parent theory.  The numerical calculation below does not evaluate
\(H^{\ret}\) or \(N\), nor does it evolve this equation in time; it studies an
effective equal-time spherical projection of its response--noise logic.
\begin{figure*}[htp]
	\centering
	\includegraphics[width=\textwidth]{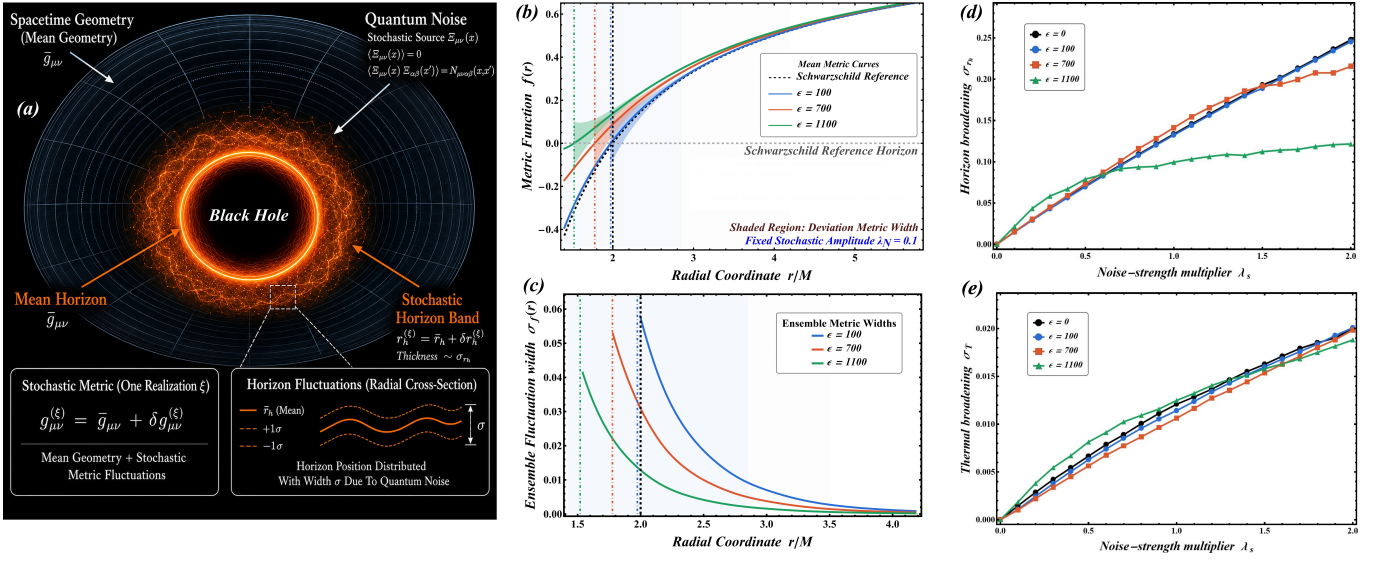}
	\caption{\label{fig:main1}
		\textbf{Effective open black-hole geometry and stochastic broadening.}
		(a) The metric is the open-system variable and quantum-field fluctuations
		are represented by an effective colored reservoir.  The horizon is reconstructed
		from every metric realization rather than displaced independently.
		(b) Mean lapses; shading is an auxiliary localization profile.
		(c) Independent covariance-propagation check of the radial metric width.
		(d) Horizon and (e) thermal broadening from the production ensemble.
		The different ordering reflects sensitivity to the value and slope of the
		same stochastic lapse.}
\end{figure*}

We use ingoing Eddington--Finkelstein coordinates so that every spherical
realization remains regular at a future horizon,
\begin{equation}
	\begin{split}
		\dd s_\xi^2={}&-e^{2\psi_\xi(v,r)}C_\xi(v,r)\dd v^2\\
		&+2e^{\psi_\xi(v,r)}\dd v\,\dd r+r^2\dd\Omega^2,\\
		C_\xi(v,r)={}&1-\frac{2Gm_\xi(v,r)}{r}.
	\end{split}
	\label{eq:metric-main}
\end{equation}
The spherical Einstein equations give the radial mass constraint
\(\partial_r m_\xi=-4\pi r^2{\cal T}^{v}{}_{v,\xi}\), the flux equation
\(\partial_v m_\xi=4\pi r^2{\cal T}^{r}{}_{v,\xi}\), and
\(\partial_r\psi_\xi=4\pi Gr\,{\cal T}_{rr,\xi}\).  We sample the ensemble in
a fixed adiabatic time window, so the slow evolution encoded by the flux
equation is frozen rather than interpreted as a simulated evaporation
history.  Accordingly, the displayed ensemble satisfies the radial mass
constraint but is not presented as a solution of the complete
time-dependent Einstein--Langevin system.

The outer trapping horizon and temperature are not supplied with independent
noise.  Both are derived from the same stochastic lapse:
\begin{equation}
	C_\xi(v,r_h^{(\xi)})=0,\qquad
	T_\xi=\left.\frac{1}{4\pi}\partial_rC_\xi(v,r)
	\right|_{r=r_h^{(\xi)}} .
	\label{eq:horizon-temperature-main}
\end{equation}
The second relation is the adiabatic Hayward--Kodama temperature of a future
outer trapping horizon \cite{Kodama1980,Hayward1998}.  Thus
\(r_h^{(\xi)}\) samples the location of a zero, while \(T_\xi\) samples the
slope at that zero.  This distinction is central to the different ordering
of the horizon and thermal widths.

\emph{Effective stochastic geometry.---}
We set \(G=\hbar=c=k_{\rm B}=1\) and \(M=1\).  The deterministic reservoir
dressing is represented by a density with an integrable exterior tail,
\begin{equation}
	\begin{aligned}
		\rho_{\eff}(r)&=\epsilon T_0^4(2M/r)^n,\qquad
		T_0=(8\pi M)^{-1},\\
		m_{\eff}(r)&=M-\frac{4\pi\epsilon T_0^4(2M)^n}{n-3}r^{3-n},\\
		f_{\eff}(r)&=1-\frac{2m_{\eff}(r)}{r}.
	\end{aligned}
	\label{eq:mean-geometry-main}
\end{equation}
The fixed-ADM-mass boundary condition is built into \(m_{\eff}\).  Here
\(T_0\) is only the Schwarzschild reference scale used to nondimensionalize
the density; it is not imposed as the temperature of a dressed branch.  The
parameter \(\epsilon\) labels an effective mean dressing, not a field
multiplicity or a microscopic coupling.

For the production ensembles,
\(\langle\xi_i(r)\xi_j(r')\rangle_{\st}
=\delta_{ij}e^{-|r-r'|/\ell_c}\), with \(\ell_c=0.18M\), and
\(\eta_i=\chi_\epsilon\xi_i\).  Here
\(\chi_\epsilon=S_w(r-\bar r_h)(\bar r_h/r)^4
e^{-\max(0,r-\bar r_h)/M}\), where
\(S_w(x)=[1+\tanh(x/w)]/2\) and \(w=0.035M\).  Only the center
\(\bar r_h(\epsilon)\) moves between branches.  A realization is
\begin{equation}
	\begin{aligned}
		\delta m_i(r)&=-4\pi\int_r^\infty s^2\eta_i(s)\dd s,\\
		f_i(r)&=f_{\eff}(r)
		-\Lambda(\lambda_s)A_{\rm src}\frac{2\delta m_i(r)}{r}.
	\end{aligned}
	\label{eq:stochastic-lapse-main}
\end{equation}
Here \(\Lambda(\lambda_s)=\lambda_s+0.20\lambda_s^2\), and the common
\(A_{\rm src}\) is fixed by
\(\sigma_f[\bar r_h(100);\lambda_s=1]=0.045\).  The same grid, fields,
covariance, and normalization are used for every \(\epsilon\).  Each retained
sample follows the zero connected to the deterministic outer branch; complete
sampling and acceptance rules are given in the Supplemental Material.

\emph{Results.---}
Figure~\ref{fig:main1}(a) summarizes the construction: the random object is
the geometry, not a horizon coordinate appended to a deterministic black
hole.  Panel (b) shows that increasing \(\epsilon\) changes the near-horizon
lapse while leaving the large-\(r\) behavior unchanged.  The outer zero moves
inward and, for the most strongly dressed branch, approaches an inner zero.
Panel (b) uses a prescribed exterior band to display the adopted
near-horizon localization; panel (c) independently checks that a localized
colored source produces a decaying exterior metric width.  Their exact,
distinct plotting prescriptions are stated in the Supplemental Material.
Quantitative horizon and thermal comparisons use the production ensemble in
Eq.~\eqref{eq:stochastic-lapse-main}; none of the profiles is claimed to be
a microscopic Unruh-state kernel.

Panels (d) and (e) are projections of the same production ensemble.
\(\sigma_{r_h}\) begins almost linearly; the \(\epsilon=1100\) curve rises
rapidly and later becomes sublinear as fluctuations probe the two-root
geometry and the acceptance condition removes invalid realizations.  At
finite strength, curvature also contains the prescribed
\(\Lambda(\lambda_s)\) and cannot be assigned to root dynamics alone.
The thermal curves cross because \(T_i\) samples both the displaced zero and
the stochastic slope.  The crossing is consequently a property of the
stated finite-noise model, not a universal ordering inherited from
\(\sigma_{r_h}\).

This difference can already be anticipated from the local geometry.  To first
order, a lapse perturbation changes the root through its value at the mean
horizon, whereas the temperature also samples the radial derivative of that
perturbation and the curvature of the mean lapse.  The horizon width is
therefore controlled mainly by a root susceptibility, while the thermal width
contains both displacement and gradient contributions.  At finite
\(\lambda_s\), evaluating these quantities at the moving root adds a nonlinear
correction.  Their finite-noise shapes combine the imposed
\(\Lambda(\lambda_s)\), evaluation at a moving root, and conditional
selection of regular future-outer realizations.

The origin of the enhanced response can be seen analytically.  For \(n=5\),
Eq.~\eqref{eq:mean-geometry-main} reduces to
\(f_{\eff}(r)=1-2/r+\epsilon/(32\pi^3r^3)\).  The simultaneous conditions
\(f_{\eff}(r_c)=f_{\eff}'(r_c)=0\) give
\begin{equation}
	r_c=\frac{4}{3},\qquad
	\epsilon_c=\frac{1024\pi^3}{27}\simeq1175.94.
	\label{eq:critical-main}
\end{equation}
The \(\epsilon=1100\) geometry is therefore subcritical but close to a double
root.  In its neighborhood,
\(r_h-r_c\propto(\epsilon_c-\epsilon)^{1/2}\) and
\(f_{\eff}'(r_h)\propto(\epsilon_c-\epsilon)^{1/2}\).  A small lapse
perturbation shifts the root according to
\begin{equation}
	\delta r_h=-\frac{\delta f(\bar r_h)}{f_{\eff}'(\bar r_h)}.
	\label{eq:susceptibility-main}
\end{equation}
The factor \(1/f_{\eff}'(\bar r_h)\) is the horizon susceptibility.  It grows
as the surface gravity decreases, even though the imposed covariance shape
and normalization are held fixed apart from translating the localization
envelope with the mean horizon.  This separates source strength from
geometric amplification and explains why \(\epsilon\) and \(\lambda_s\) play
physically different roles.

The numerical mean branches make this amplification concrete.  As
\(\epsilon\) changes from \(0\) to \(100\), \(700\), and \(1100\), the outer
root moves from \(r_h=2\) to approximately \(1.974\), \(1.776\), and \(1.520\),
while the corresponding lapse slope falls from \(0.500\) to \(0.493\),
\(0.421\), and \(0.243\).  Thus the linear root susceptibility
\(1/f_{\eff}'(r_h)\) on the strongest branch is more than twice its
Schwarzschild value even though the colored-source normalization has not been
increased.  The nearby inner root, \(r_{\rm in}\simeq1.127\) for
\(\epsilon=1100\), also explains why the finite-noise response eventually
departs from the linear prediction: sufficiently large fluctuations probe the
full two-root geometry rather than an isolated outer zero.

For each ensemble we evaluate the widths \(\sigma_{r_h}\) and \(\sigma_T\),
the covariance
\(C_{rT}=\langle(r_h^{(\xi)}-\bar r_h)(T_\xi-\bar T)\rangle_{\st}\), and the
normalized coefficient
\(\rho_{rT}=C_{rT}/(\sigma_{r_h}\sigma_T)\).  The covariance retains both the
alignment and the absolute size of the response; \(\rho_{rT}\) removes the
two widths.  At \(\lambda_s=0\), \(\rho_{rT}\) is undefined because both
widths vanish.

\begin{figure}[ht]
	\centering
	\includegraphics[width=\columnwidth]{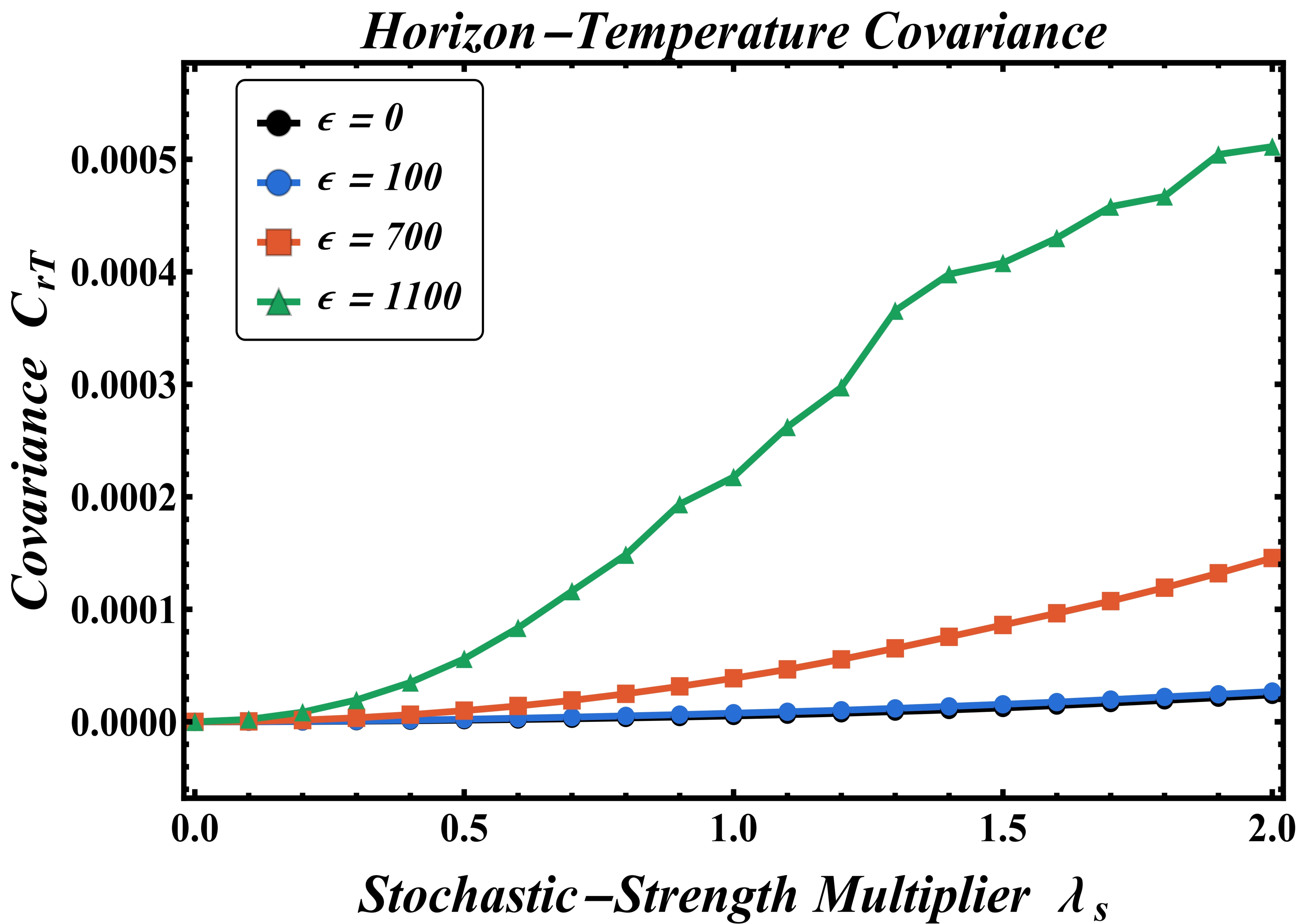}
	\caption{\label{fig:main2}
		\textbf{Horizon--temperature covariance.}
		Covariance between the reconstructed outer trapping horizon and its
		adiabatic temperature.  The enhancement for \(\epsilon=1100\) follows from
		the reduced lapse slope near the root-merger point.  Joint probability
		densities and the normalized correlation are shown in Supplemental
		Figs.~S1 and S2 \cite{Supplemental}.}
\end{figure}

Figure~\ref{fig:main2} shows that the common response of \(r_h\) and \(T\) is
weak for the Schwarzschild and \(\epsilon=100\) branches, grows for
\(\epsilon=700\), and is strongly enhanced for \(\epsilon=1100\).  The result
is not implied by either width separately.  Indeed, Fig.~\ref{fig:main1}(d)
shows that the near-critical horizon width need not be the largest at strong
noise.  Its much larger covariance instead reflects a more coherent
orientation of the joint probability cloud: fluctuations that move the root
also produce a systematic change in the slope at that root.  The linearized
relations \(\delta r_h=-\delta f_h/\bar f_h'\) and
\(\delta T=[\delta f_h'+\bar f_h''\delta r_h]/(4\pi)\) show how the
mass-like and gradient-like components of the source combine.  The full
covariance decomposition and probability densities are given in the
Supplemental Material.

The distinction between covariance and normalized correlation is important
here.  A narrow probability cloud can be well aligned while carrying little
absolute fluctuation, whereas a broad cloud can give a large covariance even
when its orientation changes only moderately.  Supplemental Fig.~S1 shows
that dressing changes both properties: the joint distribution broadens and
develops a positive tilt.  Supplemental Fig.~S2 then shows that the
coefficient \(\rho_{rT}\) rapidly approaches a branch-dependent value once
noise is present, while \(C_{rT}\) continues to grow because it retains the
two physical widths.  The pronounced green curve in Fig.~\ref{fig:main2} is
therefore not merely a normalization effect; it records the enhanced size of
the common geometric response near the root merger.

At weak stochastic strength, \(\Lambda(\lambda_s)=\lambda_s+O(\lambda_s^2)\),
so the widths begin linearly and \(C_{rT}\) begins quadratically.
At larger \(\lambda_s\), the plotted curvature combines the prescribed
amplitude map with solving for a moving zero, evaluating a derivative at that
zero, and conditioning on a regular future-outer horizon.  The robust result
is the branch-to-branch enhancement at a common imposed amplitude, not the
detailed curvature of the plotting parameter.  The smooth growth in
Fig.~\ref{fig:main2} is not a divergence or a phase transition of the source.

To examine how the geometric ensemble enters radiation, we use a scalar WKB
barrier filter computed on the dressed mean geometry of each branch.  For a
valid realization,
\begin{equation}
	L_{\gb}(T_i)=
	\sum_{\ell=0}^{\ell_{\max}}\frac{2\ell+1}{2\pi}
	\int_0^\infty\dd\omega\,
	\Gamma_{\ell}^{\rm WKB}(\omega)
	\frac{\omega}{e^{\omega/T_i}-1}.
	\label{eq:luminosity-main}
\end{equation}
The stochastic dependence in the present approximation enters through
\(T_i\); the branch-dependent barrier enters through
\(\Gamma_{\ell}^{\rm WKB}\).  We use this factor as a controlled transmission
filter, not as an exact scattering result or a fluctuating greybody
calculation.  The potentials and transmission curves are shown in
Supplemental Fig.~S3.

\begin{figure}[t]
	\centering
	\includegraphics[width=\columnwidth]{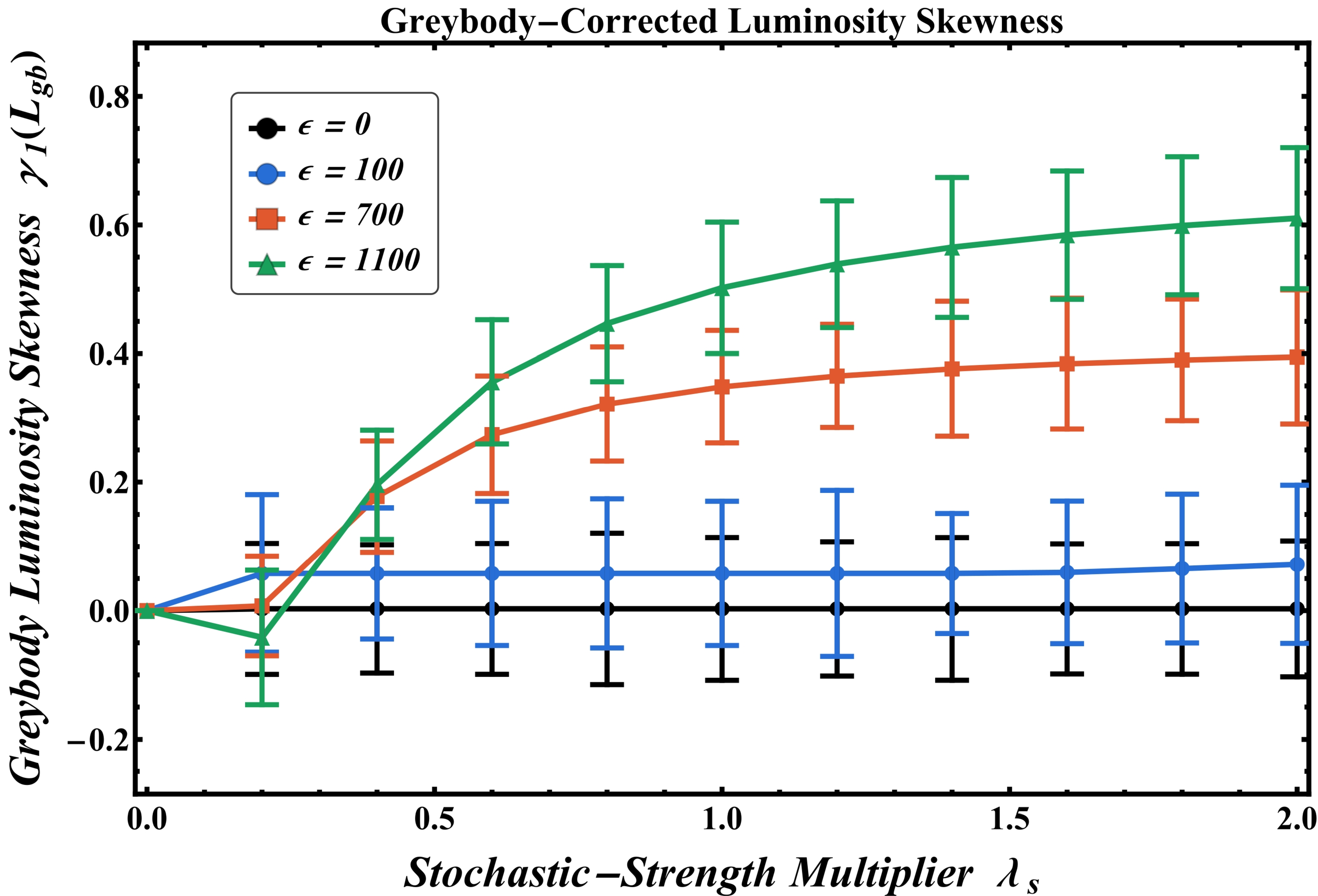}
	\caption{\label{fig:main3}
		\textbf{Radiative non-Gaussianity.}
		Corrected sample skewness of the greybody-filtered luminosity distribution.
		The weakly dressed branches remain nearly symmetric, while the
		near-critical branches develop a positive tail.  Error bars are bootstrap
		uncertainties.  At \(\lambda_s=0\) the variance vanishes and skewness is
		undefined; the plotted origin is a zero-noise convention.}
\end{figure}

Figure~\ref{fig:main3} displays the corrected skewness of \(L_{\gb}\).  A
Gaussian source does not guarantee a Gaussian radiative output: root finding,
the temperature projection, and the Planck factor are all nonlinear.  The
Schwarzschild and \(\epsilon=100\) curves remain close to zero after greybody
filtering.  The \(\epsilon=700\) and \(1100\) branches instead develop a
positive tail and approach broad plateaus.  Upward temperature excursions
are weighted more strongly than equal downward excursions by the Planck map;
near the root merger those excursions are themselves amplified by the
geometric susceptibility.  The unfiltered control in Supplemental Fig.~S2
shows that the nonlinear temperature-to-flux map already generates positive
skewness, while the WKB filter changes its magnitude and branch ordering.

More specifically, the Planck occupation factor makes \(L_{\gb}(T)\) an
increasing and locally convex function over the sampled thermal range.
Consequently, an upward fluctuation of \(T_i\) produces a larger luminosity
change than an equal downward fluctuation.  The WKB transmission does not
create this asymmetry; it reweights the frequencies over which the convex
thermal map is sampled.  The dressed barriers in Supplemental Fig.~S3
therefore alter the numerical skewness and the ordering of the branches,
whereas the unfiltered control establishes its geometric origin.  The broad
plateaus at larger \(\lambda_s\) are also natural for a standardized third
moment: once the nonlinear map controls the shape of the distribution,
increasing its overall width need not increase its skewness proportionally,
and the regular-horizon acceptance condition removes the most extreme
near-critical realizations.

\emph{Conclusion and outlook.---}
The realization-level calculation follows one prescribed colored source
through the radial constraint to joint horizon--temperature statistics and
then to a radiative observable; the auxiliary metric-band panels separately
illustrate the adopted near-horizon localization.  The common organizing
principle is the inverse-slope susceptibility near a double root.  Separating
that response from the source covariance makes the construction a benchmark
for future open-system calculations: an Unruh-state noise kernel and its
retarded partner could replace the prescribed inputs without changing the
observable projection.
This matters because the mean semiclassical equation contains no such joint
fluctuation statistics and a near-degenerate horizon can amplify modest
source correlations.  The present model remains quasistationary,
equal-time, and phenomenological, with a stochastic-semiclassical rather than
quantized metric.  It is a calculational foundation for a microscopic
open-system treatment of gravitational backreaction, not a quantum theory of
gravity.

\begin{acknowledgments}
	K.A.Y. acknowledges support from the Research Fund for International Young
	Scientists of the NSFC under Grant No.~KYZ04Y22050, Zhejiang Normal
	University research funding under Grant No.~ZC304021914, and the Zhejiang
	Province postdoctoral research project under Grant No.~ZC304021952.
\end{acknowledgments}

\paragraph{Data availability.}
The numerical data underlying Figs.~\ref{fig:main1}--\ref{fig:main3}, together
with the Mathematica routines used for sampling, root tracking,
kernel-density estimation, bootstrap analysis, and WKB filtering, are
available from the author upon reasonable request.

\end{document}


\title{Supplementary Information:\\Colored-Noise-Induced Horizon Fluctuations Near a Double Root in an Effective Black-Hole Geometry}
	\author{Kashif Ammar Yasir}
	\email{kayasir@zjnu.edu.cn}
	\affiliation{Department of Physics, Zhejiang Normal University, Jinhua 321004, China.}
	\affiliation{Zhejiang Institute of Photoelectronics, Jinhua 321004, China.}
	\setlength{\parskip}{0pt}
	\setlength{\belowcaptionskip}{-10pt}

	\date{\today}
	\maketitle
	
\section{Scope, notation, and logical status}

The Letter uses the response--noise organization of quantum Langevin theory
to construct an effective stochastic black-hole model.  This Supplemental
Material derives the formal open-system relations, the spherical projection,
and the statistical estimators, while keeping three statements with different
logical status separate.

First, the response--noise structure is general.  When environmental degrees
of freedom are eliminated from an open quantum system, their retarded
response and fluctuating force appear together in the reduced equation.
Sections~II and III derive this statement for an oscillator bath and for a
quantum field coupled to the metric.

Second, the projection from a stochastic spherical metric to its outer
trapping horizon and adiabatic temperature is geometric.  Once a realization
of the lapse is specified, its zero and slope determine \(r_h^{(\xi)}\) and
\(T_\xi\).  Sections~IV and V derive this projection and the associated
linear covariance formulas.

Third, the radial density and colored source used to produce the figures are
effective inputs.  The calculation does not evaluate a renormalized
Unruh-state stress tensor, the retarded stress-polarization kernel, or the
point-separated noise kernel.  The mean dressing is modeled by
\(\rho_{\eff}(r)\), while the stochastic sector is modeled by the envelope
\(\chi_\epsilon(r)\), the colored covariance \(K_{\ell_c}(r,r')\), and the
common normalization \(A_{\rm src}\).  Their implementation is given in
Sections~VI and VII.

Table~\ref{tab:model-map} summarizes the mapping.  References to an
environmental quantum field identify the physical motivation for the
construction; they do not imply that Unruh-state correlation functions have
been calculated here.  In particular, the Unruh state is stationary but not a
global thermal-equilibrium state, so the equilibrium
fluctuation--dissipation formula derived below cannot be used to infer its
noise kernel from a temperature alone.

\begin{table}[b]
	\caption{\label{tab:model-map}
		Formal open-system quantities and the replacements used in the numerical
		effective model.}
	\begin{ruledtabular}
		\begin{tabular}{p{0.25\textwidth}p{0.31\textwidth}p{0.34\textwidth}}
			Formal quantity & Effective realization & Status\\
			\colrule
			\(\langle\hat T_{\mu\nu}\rangle_{\rm ren}\)
			& Power-law density \(\rho_{\eff}(r)\)
			& Prescribed mean dressing\\
			\(H_{\mu\nu\alpha\beta}^{\ret}(x,x')\)
			& Quasistatic metric susceptibility
			& Exact retarded kernel not evaluated\\
			\(N_{\mu\nu\alpha\beta}(x,x')\)
			& \(\chi_\epsilon(r)K_{\ell_c}(r,r')\chi_\epsilon(r')\)
			& Phenomenological colored covariance\\
			Dynamical metric \(g_{\mu\nu}(v,r)\)
			& Equal-time ensemble \(f_i(r)\)
			& Adiabatic spherical projection\\
		\end{tabular}
	\end{ruledtabular}
\end{table}

\section{Quantum Langevin equation from an oscillator reservoir}

The analogy used in the Letter can be seen most directly in the
Caldeira--Leggett model \cite{FeynmanVernon1963,CaldeiraLeggett1983,
	BreuerPetruccione2002,Weiss2012}.  Let \(Q\) and \(P\) be a system coordinate
and momentum, and let \((q_j,p_j)\) be environmental oscillators.  A convenient
Hamiltonian, including the usual counterterm, is
\begin{equation}
	\hat H=
	\frac{\hat P^2}{2M}+V(\hat Q)
	+\sum_j\left[
	\frac{\hat p_j^2}{2m_j}
	+\frac{m_j\omega_j^2}{2}
	\left(\hat q_j-\frac{c_j\hat Q}{m_j\omega_j^2}\right)^2
	\right].
	\label{eq:CL-H}
\end{equation}
The Heisenberg equations are
\begin{align}
	M\ddot{\hat Q}(t)+V'(\hat Q(t))
	+\sum_j\frac{c_j^2}{m_j\omega_j^2}\hat Q(t)
	&=\sum_jc_j\hat q_j(t),
	\label{eq:Q-eom}\\
	\ddot{\hat q}_j(t)+\omega_j^2\hat q_j(t)
	&=\frac{c_j}{m_j}\hat Q(t).
	\label{eq:qj-eom}
\end{align}
The bath equation has the exact solution
\begin{align}
	\hat q_j(t)
	&=\hat q_j(0)\cos\omega_jt
	+\frac{\hat p_j(0)}{m_j\omega_j}\sin\omega_jt
	\nonumber\\
	&\quad
	+\frac{c_j}{m_j\omega_j}\int_0^t\dd s\,
	\sin[\omega_j(t-s)]\hat Q(s).
	\label{eq:qj-solution}
\end{align}
Substitution into Eq.~\eqref{eq:Q-eom}, followed by an integration by parts,
gives the generalized quantum Langevin equation
\begin{equation}
	M\ddot{\hat Q}(t)+V'(\hat Q(t))
	+M\int_0^t\dd s\,\gamma(t-s)\dot{\hat Q}(s)
	=\hat F(t)-M\gamma(t)\hat Q(0).
	\label{eq:QLE}
\end{equation}
The last term is the initial-slip term.  It vanishes for a suitably prepared
correlated initial state or can be absorbed into the preparation convention.
The memory kernel and fluctuating force are
\begin{align}
	\gamma(t)
	&=\frac{1}{M}\sum_j\frac{c_j^2}{m_j\omega_j^2}
	\cos\omega_jt,
	\label{eq:gamma-sum}\\
	\hat F(t)
	&=\sum_jc_j\left[
	\hat q_j(0)\cos\omega_jt
	+\frac{\hat p_j(0)}{m_j\omega_j}\sin\omega_jt
	\right].
	\label{eq:force-operator}
\end{align}
Both contain the same couplings \(c_j\).  Introducing the bath spectral
density
\begin{equation}
	J(\omega)=\frac{\pi}{2}\sum_j
	\frac{c_j^2}{m_j\omega_j}\delta(\omega-\omega_j),
	\label{eq:spectral-density}
\end{equation}
the memory kernel becomes
\begin{equation}
	\gamma(t)=\frac{2}{\pi M}\int_0^\infty\dd\omega\,
	\frac{J(\omega)}{\omega}\cos\omega t.
	\label{eq:gamma-J}
\end{equation}
For a thermal bath, the symmetrized force correlation is
\begin{align}
	\frac{1}{2}\left\langle
	\{\hat F(t),\hat F(t')\}\right\rangle
	&=
	\frac{\hbar}{\pi}\int_0^\infty\dd\omega\,J(\omega)
	\coth\!\left(\frac{\hbar\omega}{2k_{\rm B}T}\right)
	\nonumber\\
	&\hspace{1.5cm}\times\cos[\omega(t-t')].
	\label{eq:FDT}
\end{align}
Equations~\eqref{eq:gamma-J} and \eqref{eq:FDT} are the
response--fluctuation pairing referred to in the Letter.  In the Markovian
limit an Ohmic \(J(\omega)\) produces local damping and approximately white
noise.  A structured spectral density gives memory and colored noise.  The
quantum Langevin equations used in cavity optomechanics have the same
organization, although they are commonly written for damped optical and
mechanical mode operators \cite{GardinerCollett1985,Clerk2010,
	Aspelmeyer2014}.
Equation~\eqref{eq:FDT} assumes a thermal equilibrium bath.  It is included to
show why response and fluctuations must be derived from the same reservoir,
not to identify the Unruh state with a globally thermal oscillator bath.

For later comparison with the gravitational response, it is useful to expose
the susceptibility explicitly.  Linearize the system potential about a
stationary configuration, \(V''(Q_0)=M\Omega_0^2\), discard the preparation
term, and Fourier transform Eq.~\eqref{eq:QLE}.  The fluctuation obeys
\begin{equation}
	\delta\hat Q(\omega)=\chi_Q(\omega)\hat F(\omega),
	\qquad
	\chi_Q(\omega)=
	\frac{1}{
		M(\Omega_0^2-\omega^2)-iM\omega\widetilde\gamma(\omega)} .
	\label{eq:QLE-susceptibility}
\end{equation}
Consequently, its symmetrized spectrum is
\begin{equation}
	S_{QQ}(\omega)=
	|\chi_Q(\omega)|^2S_{FF}(\omega).
	\label{eq:QLE-spectrum}
\end{equation}
Equation~\eqref{eq:QLE-spectrum} separates two ingredients that must not be
confused: \(S_{FF}\) measures the noise supplied by the reservoir, whereas
\(\chi_Q\) measures how strongly the selected system responds.  A large
fluctuation of \(Q\) can therefore arise from enhanced susceptibility even
when the imposed force spectrum is unchanged.  This is precisely the
distinction between the fixed colored source and the branch-dependent
geometric amplification used in the Letter.

The gravitational correspondence is
\begin{equation}
	\hat Q\longleftrightarrow h_{\mu\nu},\qquad
	\gamma\longleftrightarrow H_{\mu\nu\alpha\beta}^{\ret},\qquad
	\hat F\longleftrightarrow\Xi_{\mu\nu}.
	\label{eq:QLE-map}
\end{equation}
This is a structural map, not an identification of the oscillator
temperature with a local black-hole temperature.  A microscopic gravitational
calculation would have to obtain \(H^{\ret}\) and \(N\) from the same quantum
field in the Unruh state.  The effective model used here instead specifies a
colored covariance and studies how the corresponding metric perturbations are
geometrically amplified.

\section{Closed-time-path origin of the Einstein--Langevin equation}

Consider a metric coupled to a quantum scalar field,
\begin{equation}
	S_{\rm tot}[g,\phi]=S_{\rm grav}[g]+S_\phi[g,\phi].
	\label{eq:S-total}
\end{equation}
The gravitational action contains the Einstein--Hilbert term and the local
curvature counterterms required by renormalization.  The environmental field
is traced out on the closed time path
\cite{MartinVerdaguer1999,PhillipsHu2001,HuVerdaguer2008,HuRoura2007}:
\begin{align}
	e^{iS_{\rm IF}[g^+,g^-]}
	&=
	\int{\cal D}\phi^+{\cal D}\phi^-\,
	\rho_{\phi}(\phi_i^+,\phi_i^-)
	\nonumber\\
	&\quad\times
	e^{\,iS_\phi[g^+,\phi^+]-iS_\phi[g^-,\phi^-]} .
	\label{eq:influence-functional}
\end{align}
The resulting effective action is
\begin{equation}
	\Gamma_{\rm CTP}[g^+,g^-]
	=S_{\rm grav}[g^+]-S_{\rm grav}[g^-]+S_{\rm IF}[g^+,g^-].
	\label{eq:CTP}
\end{equation}
Expand about a mean geometry according to
\(g_{\mu\nu}^{\pm}=\bar g_{\mu\nu}+h_{\mu\nu}^{\pm}\).  The background
\(\bar g_{\mu\nu}\) is assumed to satisfy the renormalized semiclassical mean
equation, so the terms linear in \(h^\Delta\) vanish.  Define
\(h^\Delta=h^+-h^-\) and \(h^\Sigma=(h^++h^-)/2\).  To quadratic order in
the perturbations, the nonlocal influence action has the form
\begin{align}
	S_{\rm IF}^{(2)}
	&=
	\frac{1}{2}\int\dd^4x\,\dd^4x'\,
	h_{\mu\nu}^{\Delta}(x)
	H_{\ret}^{\mu\nu\alpha\beta}(x,x')
	h_{\alpha\beta}^{\Sigma}(x')
	\nonumber\\
	&\quad
	+\frac{i}{2}\int\dd^4x\,\dd^4x'\,
	h_{\mu\nu}^{\Delta}(x)
	N^{\mu\nu\alpha\beta}(x,x')
	h_{\alpha\beta}^{\Delta}(x').
	\label{eq:influence-quadratic}
\end{align}
The real term is causal and generates the environmental response.  The
positive imaginary term suppresses widely separated histories and can be
represented by a Gaussian stochastic source.

For the connected stress operator
\begin{equation}
	\hat t_{\mu\nu}(x)=
	\hat T_{\mu\nu}(x)-\langle\hat T_{\mu\nu}(x)\rangle_{\rm ren},
	\label{eq:connected-stress}
\end{equation}
the formal noise and retarded kernels are
\begin{align}
	N_{\mu\nu\alpha\beta}(x,x')
	&=
	\frac{1}{2}\left\langle
	\{\hat t_{\mu\nu}(x),\hat t_{\alpha\beta}(x')\}
	\right\rangle,
	\label{eq:noise-kernel}\\
	H_{\mu\nu\alpha\beta}^{\ret}(x,x')
	&=
	-i\theta(t-t')
	\left\langle
	[\hat t_{\mu\nu}(x),\hat t_{\alpha\beta}(x')]
	\right\rangle
	+H_{\mu\nu\alpha\beta}^{\rm loc}(x,x').
	\label{eq:response-kernel}
\end{align}
The local term contains contact contributions and variations of the
renormalized counterterms.  Diffeomorphism invariance gives the corresponding
Ward identities, including
\(\bar\nabla^\mu N_{\mu\nu\alpha\beta}=0\) and the analogous identity for
\(H^{\ret}\), away from regulated coincidence.

Introduce a c-number Gaussian source with
\begin{equation}
	\langle\Xi_{\mu\nu}(x)\rangle_{\st}=0,\qquad
	\langle\Xi_{\mu\nu}(x)\Xi_{\alpha\beta}(x')\rangle_{\st}
	=N_{\mu\nu\alpha\beta}(x,x').
	\label{eq:Xi-cov}
\end{equation}
Varying the stochastic effective action with respect to \(h^\Delta\) and
then setting \(h^\Delta=0\) gives
\begin{equation}
	{\cal L}_{\mu\nu}{}^{\alpha\beta}h_{\alpha\beta}^{(\xi)}(x)
	+\frac{1}{2}\int\dd^4x'\,
	H_{\mu\nu}{}^{\alpha\beta,\ret}(x,x')
	h_{\alpha\beta}^{(\xi)}(x')
	=8\pi G\,\Xi_{\mu\nu}(x).
	\label{eq:EL}
\end{equation}
If \(G^{\ret}\) is the Green tensor of the complete operator on the left,
\begin{equation}
	h_{\mu\nu}^{(\xi)}(x)
	=8\pi G\int\dd^4x'\sqrt{-\bar g(x')}\,
	G_{\mu\nu}{}^{\alpha\beta,\ret}(x,x')\Xi_{\alpha\beta}(x').
	\label{eq:h-solution}
\end{equation}
Multiplying two copies of Eq.~\eqref{eq:h-solution} and using
Eq.~\eqref{eq:Xi-cov} yields
\begin{equation}
	\langle hh\rangle_{\st}
	=(8\pi G)^2G^{\ret}\star N\star G^{\adv}.
	\label{eq:metric-covariance}
\end{equation}
This is the field-theoretic counterpart of
Eqs.~\eqref{eq:gamma-J}--\eqref{eq:FDT}.  In the present numerical model,
Eq.~\eqref{eq:metric-covariance} motivates the distinction between a fixed
source covariance and a branch-dependent geometric response.  It is not
evaluated with an exact Unruh-state \(N\).  Nor are the prescribed covariance
and mean dressing obtained as a response--noise pair from one microscopic
spectral density.  The numerical construction is therefore a
Langevin-organized projection, not a first-principles solution of
Eq.~\eqref{eq:EL}.

\section{Spherical Einstein equations}

The stochastic spherical line element is
\begin{equation}
	\dd s_\xi^2
	=-e^{2\psi_\xi(v,r)}C_\xi(v,r)\dd v^2
	+2e^{\psi_\xi(v,r)}\dd v\,\dd r+r^2\dd\Omega^2,
	\qquad
	C_\xi=1-\frac{2Gm_\xi}{r}.
	\label{eq:spherical-metric}
\end{equation}
The mixed Einstein components for this parameterization can be organized as
\begin{align}
	G^{v}{}_{v}
	&=-\frac{2G}{r^2}\partial_r m_\xi,
	\label{eq:Gvv}\\
	G^{r}{}_{v}
	&=\frac{2G}{r^2}\partial_v m_\xi,
	\label{eq:Grv}\\
	G_{rr}
	&=\frac{2}{r}\partial_r\psi_\xi.
	\label{eq:Grr}
\end{align}
Using \(G_{\mu\nu}=8\pi G{\cal T}_{\mu\nu,\xi}\) gives
\begin{align}
	\partial_r m_\xi&=-4\pi r^2{\cal T}^{v}{}_{v,\xi},
	\label{eq:mass-constraint}\\
	\partial_v m_\xi&=4\pi r^2{\cal T}^{r}{}_{v,\xi},
	\label{eq:mass-flux}\\
	\partial_r\psi_\xi&=4\pi Gr\,{\cal T}_{rr,\xi}.
	\label{eq:redshift-constraint}
\end{align}
For a static density, \(\rho_\xi=-{\cal T}^{v}{}_{v,\xi}\), the first equation
becomes \(\partial_r m_\xi=4\pi r^2\rho_\xi\).  Fixing the ADM mass at infinity
therefore gives
\begin{equation}
	m_\xi(r)=M_{\rm ADM}
	-4\pi\int_r^\infty s^2\rho_\xi(s)\dd s.
	\label{eq:mass-integral-general}
\end{equation}
The sign in this expression is important: positive exterior energy reduces
the mass contained inside a finite radius while leaving the total mass at
infinity fixed.

The off-diagonal component \({\cal T}^{r}{}_{v,\xi}\) controls the slow mass
evolution.  The figures use a quasistationary snapshot and hence solve the
radial constraint at a fixed adiabatic time.  They do not integrate
Eq.~\eqref{eq:mass-flux} over an evaporation time.  Conservation,
\(\nabla_\mu{\cal T}^{\mu}{}_{\nu,\xi}=0\), is a formal requirement on a
fully time-dependent source.  The effective radial model retains the
constraint projection relevant to the displayed observables.

\section{Outer trapping horizon and temperature projection}

\subsection{Exact definitions}

For the spherical metric in Eq.~\eqref{eq:spherical-metric}, the vanishing
outgoing expansion is equivalent to
\begin{equation}
	C_\xi(v,r_h^{(\xi)})=0,
	\qquad
	r_h^{(\xi)}=2Gm_\xi(v,r_h^{(\xi)}).
	\label{eq:horizon}
\end{equation}
The intended branch is the future-outer root connected to the mean outer
solution.  Numerically, this connection is approximated by the finite-window,
nearest-root rule specified in Sec.~\ref{sec:discrete-generation} and then by the positive-temperature
cut.  The selected object is not an event horizon; it is a quasi-local
trapping-horizon candidate defined on the sampled slice.

Differentiating the first relation in Eq.~\eqref{eq:horizon} gives
\begin{equation}
	\frac{\dd r_h^{(\xi)}}{\dd v}
	=-\left.\frac{\partial_vC_\xi}{\partial_rC_\xi}
	\right|_{r=r_h^{(\xi)}}.
	\label{eq:horizon-velocity}
\end{equation}
Using \(C_\xi=1-2Gm_\xi/r\) and Eq.~\eqref{eq:mass-flux},
\begin{equation}
	\frac{\dd r_h^{(\xi)}}{\dd v}
	=
	\left.
	\frac{8\pi G r_h^{(\xi)}{\cal T}^{r}{}_{v,\xi}}
	{\partial_rC_\xi}
	\right|_{r=r_h^{(\xi)}}.
	\label{eq:horizon-flux}
\end{equation}
This relation is not used to evolve the numerical ensemble, but it shows how
the equal-time construction embeds into a dynamical one.

The Hayward--Kodama surface gravity is
\begin{equation}
	\kappa_\xi
	=\frac{1}{2}\left.
	e^{-\psi_\xi}\partial_r(e^{\psi_\xi}C_\xi)
	\right|_{r=r_h^{(\xi)}}.
	\label{eq:kappa-general}
\end{equation}
Because \(C_\xi=0\) on the horizon, the term proportional to
\(\partial_r\psi_\xi\) vanishes and
\begin{equation}
	\kappa_\xi=\frac{1}{2}
	\left.\partial_rC_\xi\right|_{r=r_h^{(\xi)}},
	\qquad
	T_\xi=\frac{\kappa_\xi}{2\pi}.
	\label{eq:kappa-temperature}
\end{equation}
The use of \(T_\xi\) assumes that the geometry varies slowly on the inverse
surface-gravity scale.  It is an instantaneous adiabatic estimator, not a
global equilibrium temperature.

\subsection{Linear root and temperature response}

Write \(C_\xi=\bar C+\delta C_\xi\) and
\(r_h^{(\xi)}=\bar r_h+\delta r_h^{(\xi)}\).  Expanding
Eq.~\eqref{eq:horizon} about \(\bar r_h\) gives
\begin{equation}
	0=
	\bar C(\bar r_h)+\bar C'_h\delta r_h^{(\xi)}
	+\delta C_\xi(\bar r_h)+O(\delta C^2).
	\label{eq:horizon-expand}
\end{equation}
Since \(\bar C(\bar r_h)=0\),
\begin{equation}
	\delta r_h^{(\xi)}
	=-\frac{\delta C_\xi(\bar r_h)}{\bar C'_h},
	\qquad
	\bar C'_h=\partial_r\bar C(\bar r_h).
	\label{eq:delta-r}
\end{equation}
This equation exposes the susceptibility \(1/\bar C'_h\).

Expanding Eq.~\eqref{eq:kappa-temperature} at the displaced root gives
\begin{align}
	T_\xi
	&=
	\frac{1}{4\pi}
	\left[
	\bar C'_h+\delta C_\xi'(\bar r_h)
	+\bar C''_h\delta r_h^{(\xi)}
	\right]+O(\delta C^2),
	\label{eq:T-expand}\\
	\delta T_\xi
	&=
	\frac{1}{4\pi}
	\left[
	\delta C_\xi'(\bar r_h)
	+\bar C''_h\delta r_h^{(\xi)}
	\right].
	\label{eq:delta-T}
\end{align}
For \(\delta C_\xi=-2G\delta m_\xi/r\),
\begin{align}
	\delta C_\xi'(\bar r_h)
	&=-\frac{2G}{\bar r_h}\partial_r\delta m_\xi(\bar r_h)
	+\frac{2G}{\bar r_h^2}\delta m_\xi(\bar r_h),
	\label{eq:delta-C-prime}\\
	\delta T_\xi
	&=
	\frac{G}{2\pi}
	\left[
	-\frac{\partial_r\delta m_\xi(\bar r_h)}{\bar r_h}
	+\frac{\delta m_\xi(\bar r_h)}{\bar r_h^2}
	+\frac{\bar C''_h\delta m_\xi(\bar r_h)}
	{\bar r_h\bar C'_h}
	\right].
	\label{eq:delta-T-mass}
\end{align}
The root is controlled by the mass perturbation at the horizon.  The
temperature contains that component, its radial gradient, and the curvature
of the mean lapse.  This is why the two widths in Fig.~1 of the Letter can
cross.

\subsection{Leading nonlinear correction}

The linear projection is sufficient to identify the susceptibility, but the
curvature of the root map becomes important for the largest values of
\(\lambda_s\).  To make this explicit, write
\(\delta r_h=\delta r_h^{(1)}+\delta r_h^{(2)}+\cdots\), where the superscript
denotes the order in the lapse perturbation.  Expanding the exact root
condition through second order gives
\begin{align}
	0={}&
	\bar C'_h(\delta r_h^{(1)}+\delta r_h^{(2)})
	+\delta C_h
	+\delta C'_h\delta r_h^{(1)}
	+\frac{1}{2}\bar C''_h(\delta r_h^{(1)})^2
	+O(\delta C^3).
	\label{eq:root-second-expand}
\end{align}
The first- and second-order displacements are therefore
\begin{align}
	\delta r_h^{(1)}
	&=-\frac{\delta C_h}{\bar C'_h},
	\label{eq:root-first}\\
	\delta r_h^{(2)}
	&=-\frac{1}{\bar C'_h}
	\left[
	\delta C'_h\delta r_h^{(1)}
	+\frac{1}{2}\bar C''_h(\delta r_h^{(1)})^2
	\right].
	\label{eq:root-second}
\end{align}
Even for a zero-mean Gaussian \(\delta C_h\), the quadratic term can generate
a nonzero displacement bias and a non-Gaussian root distribution.  More
importantly, Eqs.~\eqref{eq:root-first} and \eqref{eq:root-second} contain
increasing powers of \(1/\bar C'_h\).  Nonlinear corrections therefore become
relevant at smaller source amplitude as the mean geometry approaches a double
root.

The same expansion for the temperature yields
\begin{align}
	\delta T^{(1)}
	&=
	\frac{1}{4\pi}
	\left(\delta C'_h+\bar C''_h\delta r_h^{(1)}\right),
	\label{eq:T-first}\\
	\delta T^{(2)}
	&=
	\frac{1}{4\pi}
	\left[
	\bar C''_h\delta r_h^{(2)}
	+\delta C''_h\delta r_h^{(1)}
	+\frac{1}{2}\bar C'''_h(\delta r_h^{(1)})^2
	\right].
	\label{eq:T-second}
\end{align}
These terms explain why the finite-noise temperature width need not retain
the branch ordering predicted by the leading susceptibility alone.  They also
provide the first nonlinear stage in the conversion of a Gaussian colored
source into non-Gaussian thermal and radiative statistics.

\subsection{Covariance decomposition}

Define the equal-time mass correlators at the mean horizon by
\begin{align}
	{\cal M}_{00}
	&=\langle\delta m_h\delta m_h\rangle_{\st},\\
	{\cal M}_{01}
	&=\frac{1}{2}\left\langle
	\delta m_h\,\partial_r\delta m_h
	+\partial_r\delta m_h\,\delta m_h
	\right\rangle_{\st},\\
	{\cal M}_{11}
	&=\langle\partial_r\delta m_h\,\partial_r\delta m_h\rangle_{\st}.
	\label{eq:M-correlators}
\end{align}
Equations~\eqref{eq:delta-r} and \eqref{eq:delta-T-mass} can be written as
\begin{equation}
	\delta r_h=a_m\delta m_h,\qquad
	\delta T=A_m\delta m_h+B_m\partial_r\delta m_h,
	\label{eq:linear-projections}
\end{equation}
with
\begin{align}
	a_m&=\frac{2G}{\bar r_h\bar C'_h},\\
	A_m&=\frac{G}{2\pi}
	\left(\frac{1}{\bar r_h^2}
	+\frac{\bar C''_h}{\bar r_h\bar C'_h}\right),\\
	B_m&=-\frac{G}{2\pi\bar r_h}.
	\label{eq:projection-coefficients}
\end{align}
The linear moments are then
\begin{align}
	\sigma_{r_h}^2&=a_m^2{\cal M}_{00},
	\label{eq:var-r-linear}\\
	C_{rT}&=a_m(A_m{\cal M}_{00}+B_m{\cal M}_{01}),
	\label{eq:cov-linear}\\
	\sigma_T^2&=A_m^2{\cal M}_{00}
	+2A_mB_m{\cal M}_{01}+B_m^2{\cal M}_{11}.
	\label{eq:var-T-linear}
\end{align}
These relations explain three features of the data.  First,
\(\sigma_{r_h}\) and \(\sigma_T\) are linear in
\(\Lambda(\lambda_s)\), while \(C_{rT}\) is quadratic in
\(\Lambda(\lambda_s)\), at weak noise.  Since \(\Lambda=\lambda_s+
O(\lambda_s^2)\), the same leading powers hold in the plotted control.
Second, a small \(\bar C'_h\)
enhances both \(a_m\) and part of \(A_m\).  Third, the sign and size of
\(C_{rT}\) depend on
the balance between local-mass and mass-gradient correlations.  Covariance
therefore contains information absent from either variance separately.

\section{Effective mean geometry and its critical point}

The figures use \(G=\hbar=c=k_{\rm B}=1\), \(M=1\), and \(n=5\).  The
deterministic effective density is
\begin{equation}
	\rho_{\eff}(r)
	=\epsilon T_0^4\left(\frac{2M}{r}\right)^n,
	\qquad
	T_0=\frac{1}{8\pi M}.
	\label{eq:rho-effective}
\end{equation}
The quantity \(T_0\) is the Schwarzschild reference thermal scale used to
normalize the ansatz.  It is not imposed as the temperature of a dressed
branch.
The integral in Eq.~\eqref{eq:mass-integral-general} is
\begin{align}
	4\pi\int_r^\infty s^2\rho_{\eff}(s)\dd s
	&=
	4\pi\epsilon T_0^4(2M)^n
	\int_r^\infty s^{2-n}\dd s
	\nonumber\\
	&=
	\frac{4\pi\epsilon T_0^4(2M)^n}{n-3}r^{3-n},
	\label{eq:density-integral}
\end{align}
where \(n>3\) ensures a finite exterior tail.  Thus
\begin{align}
	m_{\eff}(r)
	&=M-\frac{4\pi\epsilon T_0^4(2M)^n}{n-3}r^{3-n},
	\label{eq:mass-effective}\\
	f_{\eff}(r)
	&=1-\frac{2m_{\eff}(r)}{r}
	\nonumber\\
	&=1-\frac{2M}{r}
	+\frac{8\pi\epsilon T_0^4(2M)^n}{n-3}r^{2-n}.
	\label{eq:lapse-effective}
\end{align}
The branches shown in the figures are
\(\epsilon=\{0,100,700,1100\}\).  The first is Schwarzschild.

For \(M=1\) and \(n=5\),
\begin{equation}
	f_{\eff}(r,\epsilon)
	=1-\frac{2}{r}+\frac{\epsilon}{32\pi^3r^3}.
	\label{eq:n5}
\end{equation}
A double root satisfies \(f_{\eff}=0\) and
\(\partial_rf_{\eff}=0\).  The derivative condition gives
\begin{equation}
	\frac{2}{r_c^2}
	-\frac{3\epsilon_c}{32\pi^3r_c^4}=0,
	\qquad
	\epsilon_c=\frac{64\pi^3}{3}r_c^2.
	\label{eq:critical-step1}
\end{equation}
Substitution in \(f_{\eff}(r_c,\epsilon_c)=0\) gives
\begin{equation}
	1-\frac{2}{r_c}+\frac{2}{3r_c}=0,
	\qquad
	r_c=\frac{4}{3},
	\label{eq:critical-step2}
\end{equation}
and therefore
\begin{equation}
	\epsilon_c=\frac{1024\pi^3}{27}\simeq1175.94.
	\label{eq:critical}
\end{equation}

The square-root scaling can also be derived explicitly.  Let
\(\delta r=r-r_c\) and \(\delta\epsilon=\epsilon-\epsilon_c\).  Because
\(f_c=f'_c=0\),
\begin{equation}
	f_{\eff}(r,\epsilon)
	\simeq
	(\partial_\epsilon f)_c\delta\epsilon
	+\frac{1}{2}(\partial_r^2f)_c\delta r^2.
	\label{eq:critical-expand}
\end{equation}
At the critical point,
\((\partial_\epsilon f)_c=27/(2048\pi^3)\) and
\((\partial_r^2f)_c=27/16\).  Hence
\begin{equation}
	\delta r_{\rm out/in}
	=\pm\sqrt{\frac{\epsilon_c-\epsilon}{64\pi^3}}
	+O(\epsilon_c-\epsilon).
	\label{eq:root-splitting}
\end{equation}
The slope of the outer branch is
\begin{equation}
	f_{\eff}'(r_{\rm out})
	\simeq\frac{27}{128\pi^{3/2}}
	\sqrt{\epsilon_c-\epsilon}.
	\label{eq:slope-scaling}
\end{equation}
Equations~\eqref{eq:root-splitting} and \eqref{eq:slope-scaling} make the
critical amplification quantitative.  The \(\epsilon=1100\) branch lies
below \(\epsilon_c\), so it still has a regular outer root, but its
susceptibility \(1/f_{\eff}'(r_h)\) is much larger than on the other branches.

\begin{table}[t]
	\caption{\label{tab:mean-branches}
		Mean-geometry quantities for the branches used in the figures.  The
		temperature is \(T_{\eff}=f_{\eff}'(r_{\rm out})/(4\pi)\), and
		\(\mathcal S_h=1/f_{\eff}'(r_{\rm out})\) is the linear root
		susceptibility.}
	\begin{ruledtabular}
		\begin{tabular}{cccccc}
			\(\epsilon\) & \(r_{\rm in}\) & \(r_{\rm out}\) &
			\(f_{\eff}'(r_{\rm out})\) & \(T_{\eff}\) & \(\mathcal S_h\)\\
			\colrule
			\(0\)    & ---     & \(2.000\) & \(0.500\) & \(0.03979\) & \(2.000\)\\
			\(100\)  & \(0.239\) & \(1.974\) & \(0.493\) & \(0.03925\) & \(2.027\)\\
			\(700\)  & \(0.752\) & \(1.776\) & \(0.421\) & \(0.03352\) & \(2.374\)\\
			\(1100\) & \(1.127\) & \(1.520\) & \(0.243\) & \(0.01932\) & \(4.119\)\\
		\end{tabular}
	\end{ruledtabular}
\end{table}

Table~\ref{tab:mean-branches} shows that the strongly dressed branch is
qualitatively different even before noise is added.  Its inner and outer roots
are separated by only \(0.394M\), and its lapse slope is less than one half of
the Schwarzschild value.  The leading critical estimate,
\(r_{\rm out/in}\simeq r_c\pm
\sqrt{(\epsilon_c-\epsilon)/(64\pi^3)}\), gives \(1.529\) and \(1.138\) for
\(\epsilon=1100\), close to the exact numerical roots in the table.  This
agreement confirms that the enhanced response is controlled by the
root-merger asymptotics rather than by an accidental feature of the sampling.

\section{Colored source and numerical realization}

\subsection{Production kernel, envelope, and normalization}

The realization-level horizon and temperature calculations use independent
zero-mean, unit-variance Gaussian radial fields with exponential covariance,
\begin{equation}
	\langle\xi_i(r)\xi_j(r')\rangle_{\st}
	=\delta_{ij}K_{\ell_c}(r,r'),\qquad
	K_{\ell_c}(r,r')=
	\exp\!\left(-\frac{|r-r'|}{\ell_c}\right),
	\quad \ell_c=0.18M .
	\label{eq:colored-covariance}
\end{equation}
This is a spatial Ornstein--Uhlenbeck covariance; it is an effective
equal-time prescription, not a noise kernel derived from the Unruh state.
Define
\begin{equation}
	S_w(x)=\frac{1}{2}\left[1+\tanh\!\left(\frac{x}{w}\right)\right],
	\qquad d_\epsilon(r)=\max[0,r-\bar r_h(\epsilon)] ,
	\label{eq:smooth-switch}
\end{equation}
with \(w=0.035M\).  The source envelope used in the production ensemble is
\begin{equation}
	\chi_\epsilon(r)=
	S_w[r-\bar r_h(\epsilon)]
	\left[\frac{\bar r_h(\epsilon)}{r}\right]^4
	\exp\!\left[-\frac{d_\epsilon(r)}{M}\right],
	\qquad
	\eta_i(r)=\chi_\epsilon(r)\xi_i(r).
	\label{eq:eta}
\end{equation}
Thus the correlation length, switch width, power, and exterior decay length
are the same for all branches; only the center \(\bar r_h(\epsilon)\) moves.
The source covariance is consequently
\begin{equation}
	\langle\eta_i(r)\eta_j(r')\rangle_{\st}
	=\delta_{ij}\chi_\epsilon(r)
	K_{\ell_c}(r,r')\chi_\epsilon(r').
	\label{eq:eta-covariance}
\end{equation}
In units with \(M=1\), \(\xi\) and \(\chi_\epsilon\) are dimensionless and
the common factor \(A_{\rm src}\) introduced below supplies the density
dimension \(M^{-2}\).

The boundary condition \(\delta m_i(\infty)=0\) gives
\begin{align}
	\delta m_i(r)
	&=-4\pi\int_r^\infty s^2\eta_i(s)\dd s,
	\label{eq:delta-m}\\
	\delta f_i(r)
	&=-\frac{2\delta m_i(r)}{r}.
	\label{eq:delta-f}
\end{align}
Differentiating Eq.~\eqref{eq:delta-m} gives
\(\delta m_i'(r)=4\pi r^2\eta_i(r)\), so
\begin{equation}
	\delta f_i'(r)
	=-8\pi r\eta_i(r)+\frac{2\delta m_i(r)}{r^2}.
	\label{eq:delta-f-prime}
\end{equation}
The full sampled lapse and slope are
\begin{align}
	f_i(r;\epsilon,\lambda_s)
	&=f_{\eff}(r;\epsilon)
	+\Lambda(\lambda_s)A_{\rm src}\delta f_i(r;\epsilon),
	\label{eq:full-lapse}\\
	f_i'(r;\epsilon,\lambda_s)
	&=f_{\eff}'(r;\epsilon)
	+\Lambda(\lambda_s)A_{\rm src}\delta f_i'(r;\epsilon),
	\label{eq:full-slope}\\
	\Lambda(\lambda_s)&=\lambda_s+0.20\lambda_s^2 .
	\label{eq:lambda-map}
\end{align}
The control values are
\(\lambda_s=0,0.1,\ldots,2.0\).  Equation~\eqref{eq:lambda-map} is the
finite-strength amplitude map present in the production code.  Therefore,
curvature of an observable plotted against \(\lambda_s\) contains this
prescribed reparametrization as well as nonlinear geometric response.  The
near-zero response is unaffected because
\(\Lambda(\lambda_s)=\lambda_s+O(\lambda_s^2)\).

The common source scale is not adjusted independently for each branch.  It
is fixed once on the \(\epsilon_{\rm ref}=100\) branch by
\begin{equation}
	A_{\rm src}=
	\frac{0.045}
	{\Lambda(1)\,\widehat{\sigma}^{\,({\rm unit})}_f
		[\bar r_h(100)]},
	\qquad \Lambda(1)=1.20 ,
	\label{eq:source-normalization}
\end{equation}
where \(\widehat{\sigma}^{\,({\rm unit})}_f\) is the standard deviation of
the unit-amplitude lapse response evaluated at the grid point nearest the
mean outer horizon, using the same 650 seeded realizations.  The target is
therefore
\(\sigma_f[\bar r_h(100);\lambda_s=1]=0.045\).  Direct evaluation of the
discrete covariance gives
\(\sigma_f^{({\rm unit})}\simeq12.846M^2\), corresponding to the
ensemble-limit estimate \(A_{\rm src}\simeq2.92\times10^{-3}M^{-2}\).
The implementation uses Eq.~\eqref{eq:source-normalization}, so its
finite-sample value is fixed by the stated realization set rather than by
this rounded estimate.

\subsection{Auxiliary profiles in Fig.~1(b,c)}

The shaded profile in Fig.~1(b) of the Letter is a visualization aid rather
than a second sample of the production ensemble.  The plotting code prescribes
\begin{equation}
	\sigma_f^{({\rm vis})}(r;\epsilon)
	=0.10\,\Theta[r-\bar r_h(\epsilon)]
	\left[\frac{\bar r_h(\epsilon)}{r}\right]^4
	\exp\!\left[-\frac{r-\bar r_h(\epsilon)}{M}\right]
	\label{eq:visual-band}
\end{equation}
for \(\epsilon=100,700,\) and \(1100\).  It is used only to show where a
horizon-localized metric band would lie around the exact mean lapse; it does
not enter the horizon, temperature, covariance, or luminosity statistics.

Figure~1(c) is an independent covariance-propagation check.  That script uses
the squared-exponential kernel
\begin{equation}
	K^{({\rm G})}_{\ell_c}(r,r')
	=\exp\!\left[-\frac{(r-r')^2}{2\ell_c^2}\right],
	\qquad \ell_c=0.18M ,
	\label{eq:gaussian-check-kernel}
\end{equation}
and the sharp exterior envelope
\(\Theta[r-\bar r_h](\bar r_h/r)^4
\exp[-(r-\bar r_h)/M]\).  It uses 280 points on
\(r/M\in[1.30,8.00]\), \(\Delta r\simeq0.0240143M\), 800 realizations,
seed 123456, and a Cholesky factorization of the covariance after adding
\(10^{-9}\) to its diagonal.  Its common amplitude is calibrated by
\(\sigma_f[\bar r_h(100)]=0.060\); the corresponding covariance-limit
estimate is \(A_{\rm src}^{({\rm G})}\simeq4.05\times10^{-3}M^{-2}\).
The script evaluates convergence at \(N=200,400,\) and 800.  These settings
explain why panel (c) should be read as a localization and convergence check,
not as a quantitative slice through the production ensemble of panels
1(d,e).

\subsection{From the colored kernel to metric covariance}

The effective covariance can be propagated analytically through the radial
mass constraint.  Define the scaled stochastic mass perturbation by
\(\Delta m_i(r)=\Lambda(\lambda_s)A_{\rm src}\delta m_i(r)\).  Substitution of
Eq.~\eqref{eq:delta-m} into Eq.~\eqref{eq:eta-covariance} gives
\begin{align}
	\left\langle\Delta m(r)\Delta m(r')\right\rangle_{\st}
	&=[\Lambda(\lambda_s)A_{\rm src}]^2{\cal M}_{\epsilon}(r,r'),
	\label{eq:mass-covariance-colored}\\
	{\cal M}_{\epsilon}(r,r')
	&=(4\pi)^2
	\int_r^\infty\dd s\,s^2
	\int_{r'}^\infty\dd s'\,{s'}^2
	\chi_\epsilon(s)K_{\ell_c}(s,s')\chi_\epsilon(s').
	\label{eq:M-kernel}
\end{align}
Thus the mass constraint performs two radial integrations of the colored
source.  This smoothing is why the sampled lapse can remain differentiable
even when the source varies on the shorter scale \(\ell_c\).

Since \(\Delta f(r)=-2\Delta m(r)/r\), the lapse covariance is
\begin{equation}
	{\cal C}_f(r,r')
	\equiv
	\left\langle\Delta f(r)\Delta f(r')\right\rangle_{\st}
	=
	\frac{4[\Lambda(\lambda_s)A_{\rm src}]^2}{rr'}
	{\cal M}_{\epsilon}(r,r').
	\label{eq:lapse-covariance-colored}
\end{equation}
Correlators involving the stochastic slope follow by differentiating the
appropriate argument of \({\cal C}_f\):
\begin{align}
	\left\langle\Delta f'(r)\Delta f(r')\right\rangle_{\st}
	&=\partial_r{\cal C}_f(r,r'),\\
	\left\langle\Delta f'(r)\Delta f'(r')\right\rangle_{\st}
	&=\partial_r\partial_{r'}{\cal C}_f(r,r').
\end{align}
Introduce the local projection operator
\begin{equation}
	{\cal D}_h=\left.
	\left(\partial_r-\frac{\bar f_h''}{\bar f_h'}\right)
	\right|_{r=\bar r_h}.
	\label{eq:projection-operator}
\end{equation}
Below, \({\cal D}_r\) and \({\cal D}_{r'}\) denote this operator acting on the
first and second arguments, respectively, before both are set to
\(\bar r_h\).
The first-order relations can then be written as
\(\delta r_h=-\Delta f_h/\bar f_h'\) and
\(\delta T={\cal D}_h\Delta f/(4\pi)\).  Directly in terms of
\({\cal C}_f\),
\begin{align}
	\sigma_{r_h}^2
	&=\frac{{\cal C}_f(\bar r_h,\bar r_h)}{(\bar f_h')^2},
	\label{eq:sigma-r-from-Cf}\\
	C_{rT}
	&=-\frac{1}{4\pi\bar f_h'}
	\left.
	{\cal D}_{r'}{\cal C}_f(r,r')
	\right|_{r=r'=\bar r_h},
	\label{eq:CrT-from-Cf}\\
	\sigma_T^2
	&=\frac{1}{(4\pi)^2}
	\left.
	{\cal D}_{r}{\cal D}_{r'}{\cal C}_f(r,r')
	\right|_{r=r'=\bar r_h}.
	\label{eq:sigma-T-from-Cf}
\end{align}
Equations~\eqref{eq:M-kernel}--\eqref{eq:sigma-T-from-Cf} give the complete
linear chain from the prescribed colored covariance to the three geometric
statistics displayed in the Letter.  They also show explicitly that every
linearized variance contains the common factor
\([\Lambda(\lambda_s)A_{\rm src}]^2\), whereas the strong branch dependence
enters through \(\bar r_h\), \(\bar f_h'\), and \(\bar f_h''\).

\subsection{Discrete generation of colored samples}
\label{sec:discrete-generation}

The production grid contains \(N_r=680\) uniformly spaced points on
\(r/M\in[1,9]\), so
\(\Delta r=8M/679\simeq0.0117820M\).  With
\(q=\exp(-\Delta r/\ell_c)\) and independent
\(z_{i,a}\sim{\cal N}(0,1)\), the field is generated recursively as
\begin{equation}
	\xi_{i,1}=z_{i,1},\qquad
	\xi_{i,a}=q\xi_{i,a-1}+\sqrt{1-q^2}\,z_{i,a}
	\quad(a\geq2).
	\label{eq:colored-discrete}
\end{equation}
It follows directly that
\(\langle\xi_{i,a}\xi_{j,b}\rangle_{\st}
=\delta_{ij}q^{|a-b|}
=\delta_{ij}\exp[-|r_a-r_b|/\ell_c]\).  The calculation uses
\(N_{\rm gen}=650\) common field realizations for every branch and every
\(\lambda_s\), initialized with the pseudorandom seed 12345.  Reusing the
same \(z_{i,a}\) across branches is a variance-reduction device for comparing
their response; it does not correlate distinct members \(i\) of an
individual ensemble.

Multiplication by \(\chi_\epsilon(r_a)\) gives \(\eta_{i,a}\).
Equation~\eqref{eq:delta-m} is evaluated from the outer boundary inward using
a cumulative trapezoidal rule, with
\(\delta m_i(r_{\max})=0\).  Thus the continuum condition at infinity is
implemented by truncating the exponentially localized source at
\(r_{\max}=9M\).  The mean horizons are
\(\bar r_h/M=2,1.974139,1.776438,\) and \(1.520405\) for
\(\epsilon=0,100,700,\) and \(1100\), respectively, all well inside the
grid.

For each \((\epsilon,\lambda_s)\), the numerical sequence is:
\begin{enumerate}
	\item construct \(f_i(r_a;\epsilon,\lambda_s)\) and its slope;
	\item identify all zero crossings in the physical radial interval;
	\item retain candidate zeros in
	\([\bar r_h-0.65M,\bar r_h+0.90M]\), linearly interpolate each
	sign-changing pair, and select the candidate closest to \(\bar r_h\);
	\item evaluate the stochastic source and mass at the selected zero using,
	respectively, linear and quadratic interpolation, and then evaluate
	\(f_i'\) from Eq.~\eqref{eq:delta-f-prime};
	\item set \(T_i=f_i'(r_{h,i})/(4\pi)\);
	\item retain the realization only if the tracked root exists and
	\(0<T_i\leq0.25M^{-1}\).
\end{enumerate}
The positive-temperature requirement selects a future-outer candidate.  The
upper temperature bound is an additional numerical admissibility cutoff in
the supplied implementation; it is not derived from the mean geometry.
Only parameter points with \(N_{\rm valid}\geq140\) are plotted.  The final
ensemble is therefore conditional on the root window and temperature cuts.
Near \(\epsilon_c\), large perturbations can merge or exchange roots, so the
retained count \(N_{\rm valid}\) and
\(p_{\rm acc}=N_{\rm valid}/650\) are part of the numerical result rather
than optional implementation details.


\section{Statistics and density estimation}

For \(N_{\rm valid}\) retained samples, the means and unbiased sample
variances are
\begin{align}
	\bar r_h&=\frac{1}{N_{\rm valid}}\sum_i r_{h,i},&
	\sigma_{r_h}^2&=\frac{1}{N_{\rm valid}-1}
	\sum_i(r_{h,i}-\bar r_h)^2,
	\label{eq:r-stats}\\
	\bar T&=\frac{1}{N_{\rm valid}}\sum_i T_i,&
	\sigma_T^2&=\frac{1}{N_{\rm valid}-1}
	\sum_i(T_i-\bar T)^2.
	\label{eq:T-stats}
\end{align}
The covariance and normalized coefficient are
\begin{align}
	C_{rT}
	&=\frac{1}{N_{\rm valid}-1}
	\sum_i(r_{h,i}-\bar r_h)(T_i-\bar T),
	\label{eq:sample-covariance}\\
	\rho_{rT}
	&=\frac{C_{rT}}{\sigma_{r_h}\sigma_T}.
	\label{eq:sample-correlation}
\end{align}
At \(\lambda_s=0\), both widths vanish and \(\rho_{rT}\) is undefined.
The origin shown in Supplemental Fig.~\ref{fig:S2-correlation}(a) is only a
plotting convention.

\begin{figure}[t]
	\centering
	\includegraphics[width=0.72\textwidth]{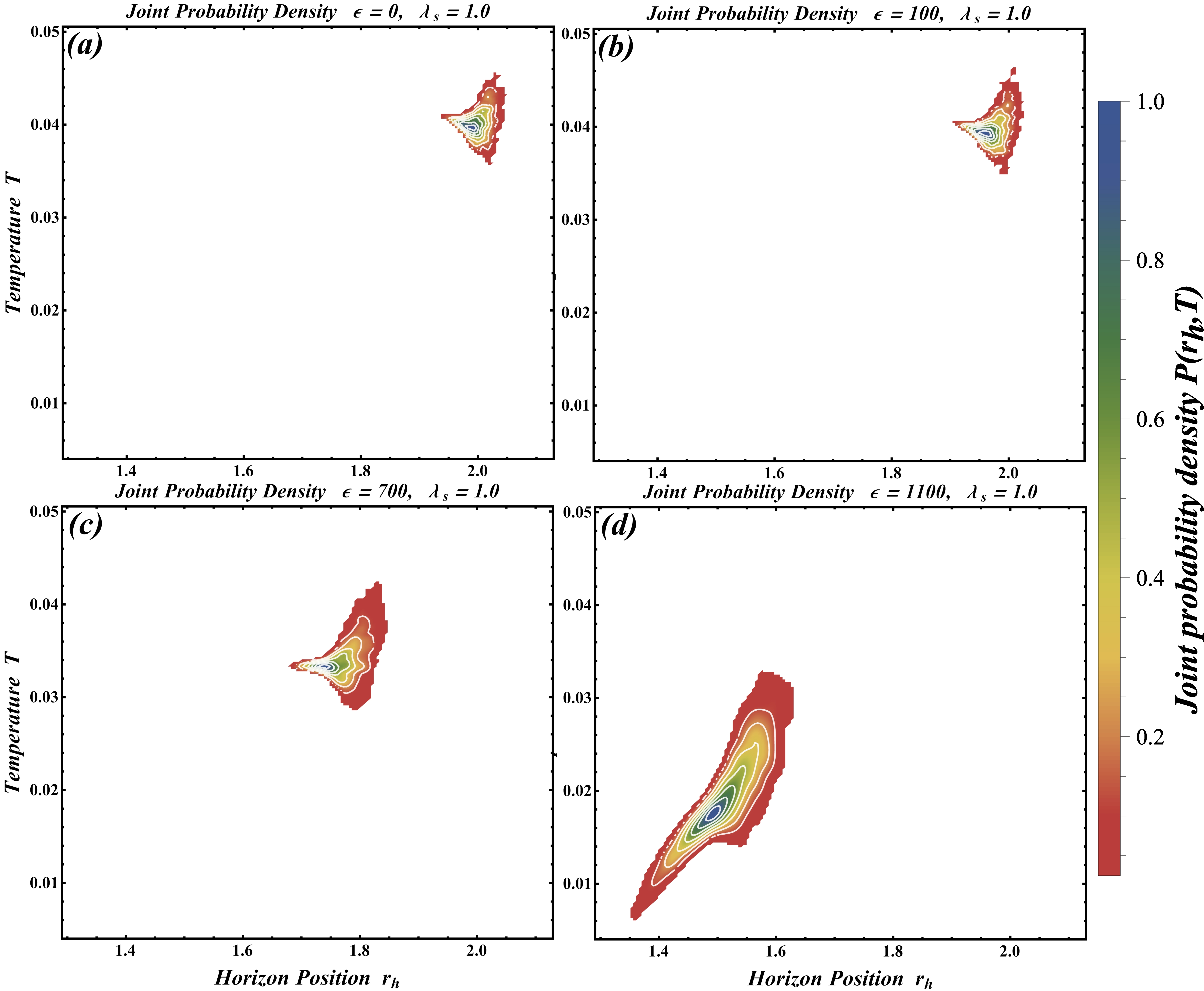}
	\caption{\label{fig:S1-joint}
		\textbf{Joint horizon--temperature probability density.}
		Kernel-density estimates at \(\lambda_s=1\) for
		\(\epsilon=0,100,700,\) and \(1100\).  Weakly dressed branches remain
		compact.  Strong dressing shifts and broadens the distribution; the
		near-critical branch develops a pronounced positive tilt.  Each panel is
		normalized to its own maximum.}
\end{figure}

The joint distribution is
\begin{equation}
	P_{\epsilon,\lambda_s}(r_h,T)
	=\left\langle
	\delta(r_h-r_h^{(\xi)})
	\delta(T-T_\xi)
	\right\rangle_{\st}.
	\label{eq:joint-formal}
\end{equation}
The displayed Gaussian kernel-density estimator is
\begin{equation}
	\widehat P(r_h,T)
	=\frac{1}{2\pi N_{\rm valid}h_r h_T}
	\sum_i
	\exp\!\left[
	-\frac{(r_h-r_{h,i})^2}{2h_r^2}
	-\frac{(T-T_i)^2}{2h_T^2}
	\right].
	\label{eq:KDE}
\end{equation}
Each panel in Fig.~\ref{fig:S1-joint} is divided by its own maximum.  Color
therefore compares shape within a panel, not absolute probability density
between different branches.

\section{Detailed interpretation of the geometric results}

\subsection{Mean lapse and metric width}

Figure~1(b) of the Letter compares \(f_{\eff}(r)\) for the four values of
\(\epsilon\).  Because the ADM mass is fixed, all curves approach
\(1-2M/r\) at large radius.  Their separation is confined to the region in
which the power-law tail in Eq.~\eqref{eq:rho-effective} is appreciable.
Increasing \(\epsilon\) raises the positive \(r^{-3}\) term in
Eq.~\eqref{eq:n5}; the outer root therefore moves inward from the
Schwarzschild value \(r_h=2\).  On the \(\epsilon=1100\) branch, the outer
and inner roots are already close compared with the weakly dressed branches.

The shading in Fig.~1(b) is the prescribed profile
Eq.~\eqref{eq:visual-band}; it is not used in subsequent ensemble
statistics.  Figure~1(c) shows the lapse standard deviation obtained in the
separate squared-exponential-kernel check of
Eq.~\eqref{eq:gaussian-check-kernel}.  The mass integral smooths that source,
while division by \(r\) converts it to \(\delta f\).  The peak follows the
mean horizon and the width decays outside the source region, confirming the
intended localization.  Because panels (b) and (c) are auxiliary
visualization checks, their amplitudes and kernel should not be numerically
identified with the production ensemble or with a microscopic Unruh-state
noise kernel.

\subsection{Horizon and thermal widths}

In the linear regime, Eq.~\eqref{eq:full-lapse} implies
\(\delta f\propto\Lambda(\lambda_s)\).  Since
\(\Lambda(\lambda_s)=\lambda_s+O(\lambda_s^2)\),
\(\sigma_{r_h}\) and \(\sigma_T\) begin linearly in \(\lambda_s\).  This
accounts for the nearly linear onset in Figs.~1(d) and 1(e) of the Letter.
The slopes differ among
branches because the projection coefficients in
Eq.~\eqref{eq:projection-coefficients} depend on
\(\bar C'_h\), \(\bar C''_h\), and \(\bar r_h\).

The sublinear horizon broadening of the \(\epsilon=1100\) curve at larger
\(\lambda_s\) does not contradict its large linear susceptibility.  The
linear expression in Eq.~\eqref{eq:delta-r} is valid only while the root
remains close to \(\bar r_h\).  At finite noise, the prescribed amplitude map,
the curvature of the lapse, the nearby inner root, and the conditional
selection rule all affect the plotted shape.  Realizations that leave the
tracking window or fail the temperature cuts do not contribute.  The robust
comparison is therefore the enhanced initial response at common imposed
amplitude, not a universal interpretation of the large-\(\lambda_s\)
plateau.

The thermal curves cross because Eq.~\eqref{eq:delta-T-mass} contains both
\(\delta m_h\) and \(\partial_r\delta m_h\).  A branch with a large root
susceptibility does not necessarily maximize the gradient variance.  The
curvature term can also reinforce or partially cancel the direct stochastic
slope.  Thus \(\sigma_T\) and \(\sigma_{r_h}\) are genuinely different
observables rather than rescaled versions of one another.

\subsection{Joint density and covariance}

Supplemental Fig.~\ref{fig:S1-joint} makes the covariance visible at the
distribution level.  For \(\epsilon=0\) and \(100\), the clouds are compact
and only weakly tilted.  The \(\epsilon=700\) cloud broadens and shifts, while
the \(\epsilon=1100\) cloud becomes elongated along a positive-slope
direction.  A positive displacement of the root is then statistically
associated with a positive change of the temperature estimator.  The tilt,
not broadening alone, produces the large \(C_{rT}\) in Fig.~2 of the Letter.

\begin{figure}[t]
	\centering
	\includegraphics[width=0.70\textwidth]{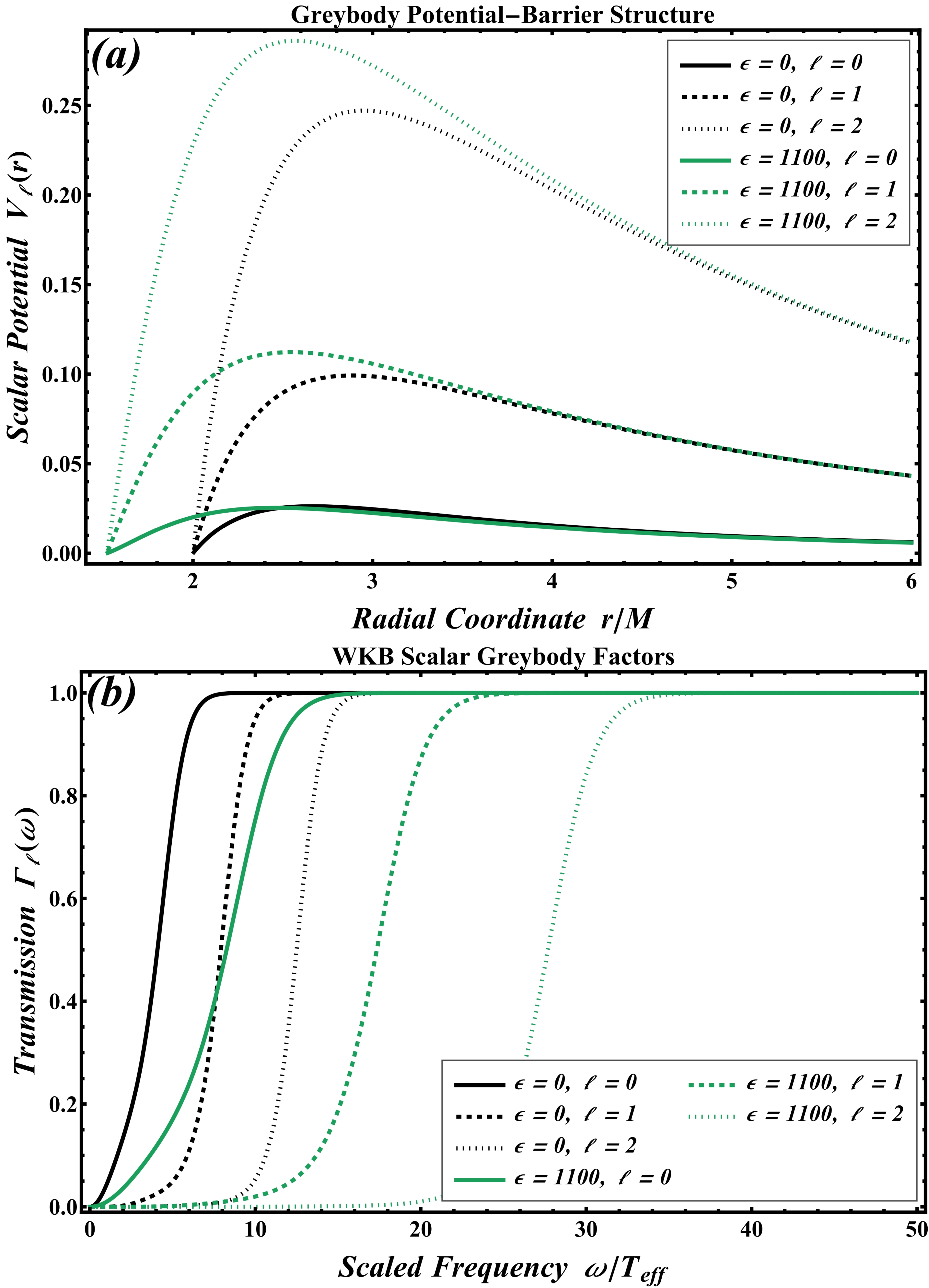}
	\caption{\label{fig:S2-correlation}
		\textbf{Normalized correlation and unfiltered luminosity control.}
		(a) Normalized horizon--temperature correlation coefficient.  At zero
		noise the coefficient is undefined; the displayed origin is a plotting
		convention.  (b) Skewness of the effective luminosity before greybody
		filtering.  The control demonstrates that the nonlinear
		horizon--temperature-to-luminosity map already produces positive tails.}
\end{figure}

Supplemental Fig.~\ref{fig:S2-correlation}(a) separates alignment from
absolute scale.  Once nonzero noise is present, \(\rho_{rT}\) rises rapidly
and then varies slowly because the common factor
\(\Lambda(\lambda_s)^2\) cancels
between covariance and widths.  The \(\epsilon=1100\) branch remains most
strongly aligned, the \(\epsilon=700\) branch is intermediate, and the weakly
dressed branches are much less correlated.  By contrast, the covariance in
the Letter continues to grow because it retains the absolute fluctuation
amplitudes.  Reporting both quantities prevents a large correlation
coefficient from being confused with a large physical response.

Equation~\eqref{eq:CrT-from-Cf} makes this distinction quantitative.  Written
out at the mean horizon,
\begin{equation}
	C_{rT}=
	-\frac{1}{4\pi\bar f_h'}
	\left[
	\partial_{r'}{\cal C}_f(\bar r_h,r')\big|_{r'=\bar r_h}
	-\frac{\bar f_h''}{\bar f_h'}
	{\cal C}_f(\bar r_h,\bar r_h)
	\right].
	\label{eq:CrT-explicit-discussion}
\end{equation}
The first term correlates the lapse value with its stochastic gradient.  The
second is the root-displacement contribution and contains two inverse powers
of the mean slope.  Near the merger, \(\bar f_h'\) decreases while
\(\bar f_h''\) remains finite, so the covariance can be amplified much more
strongly than the underlying colored covariance.  Its sign still depends on
the balance between the two terms; the observed positive tilt shows that
their combination is positive for the chosen kernel and envelope.  This is
why the result is a statement about the response of this effective geometry,
not a universal assertion that every stochastic horizon must have positive
horizon--temperature covariance.

\section{Greybody filter and radiative statistics}

\subsection{Scalar potential}

For a static spherical mean geometry,
\begin{equation}
	\dd s^2=-A(r)\dd t^2+B(r)\dd r^2+r^2\dd\Omega^2,
	\label{eq:static-metric}
\end{equation}
introduce the tortoise coordinate
\(\dd r_*/\dd r=\sqrt{B/A}\).  After separating a scalar field into
spherical harmonics, the radial mode obeys
\begin{equation}
	\left[
	\frac{\dd^2}{\dd r_*^2}+\omega^2-V_\ell(r)
	\right]\psi_{\omega\ell}(r)=0.
	\label{eq:radial-wave}
\end{equation}
For scalar mass \(m_\phi\) and curvature coupling \(\xi_\phi\),
\begin{equation}
	V_\ell(r)
	=A(r)\left[
	\frac{\ell(\ell+1)}{r^2}+m_\phi^2+\xi_\phi R
	+\frac{1}{r\sqrt{AB}}\frac{\dd}{\dd r}
	\sqrt{\frac{A}{B}}
	\right].
	\label{eq:potential}
\end{equation}
For the lapse representation used in the figures,
\(A=f_{\eff}\) and \(B=f_{\eff}^{-1}\), so the last term reduces to
\(f_{\eff}'/r\).  The plotted barriers and luminosity use a massless,
minimally coupled scalar, \(m_\phi=\xi_\phi=0\).

Below the barrier maximum, the WKB transmission is
\begin{equation}
	\Gamma_\ell^{\rm WKB}(\omega)
	\simeq
	\left[
	1+\exp\!\left(
	2\int_{r_{*,1}}^{r_{*,2}}
	\sqrt{V_\ell(r_*)-\omega^2}\,\dd r_*
	\right)
	\right]^{-1},
	\label{eq:WKB}
\end{equation}
where the turning points satisfy \(V_\ell=\omega^2\).  Above the barrier the
interpolation tends smoothly to unity.  The approximation captures how the
mean curvature barrier reweights different frequencies; it is not a
replacement for numerical scattering with exact boundary conditions.

\begin{figure}[t]
	\centering
	\includegraphics[width=0.63\textwidth]{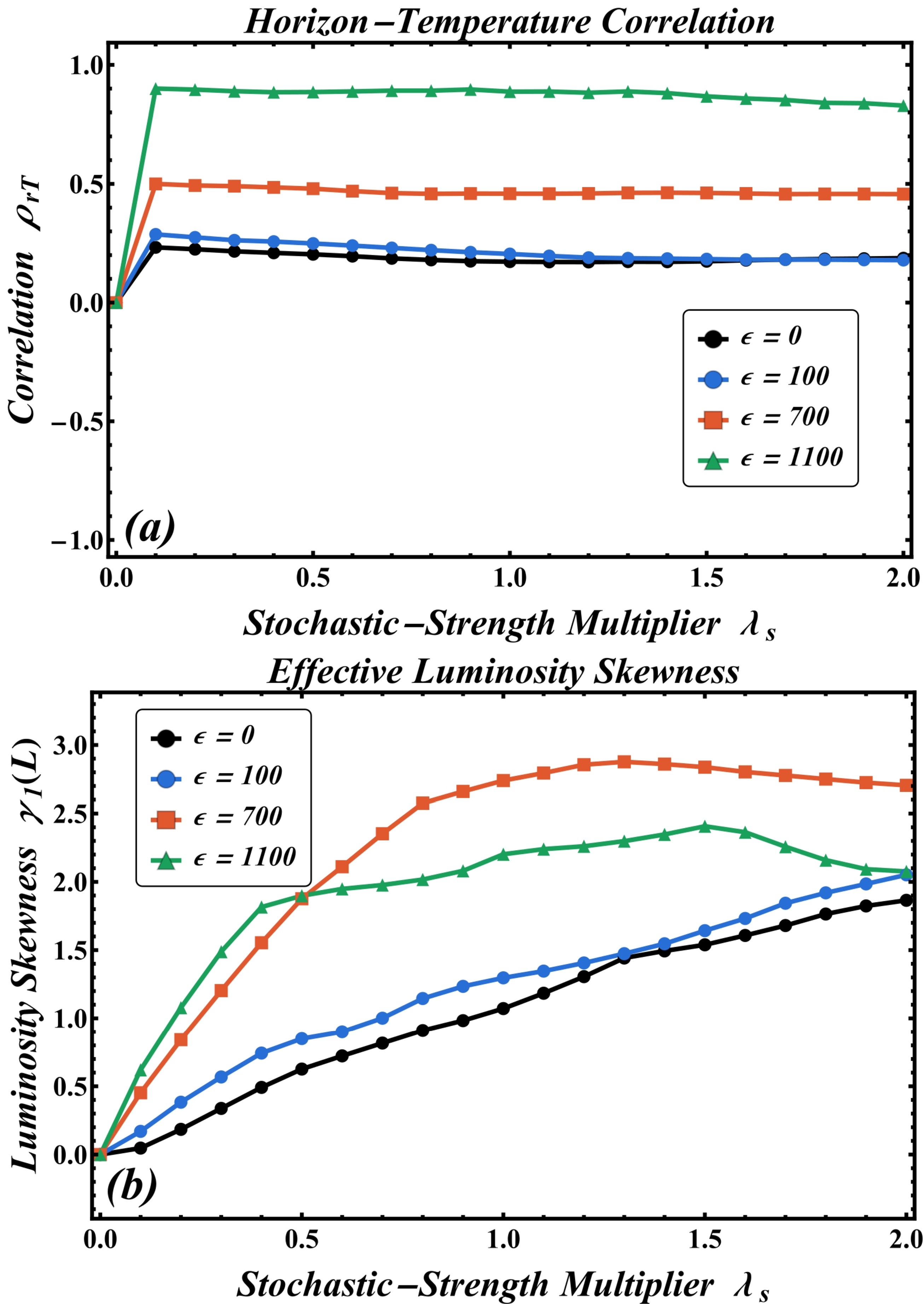}
	\caption{\label{fig:S3-greybody}
		\textbf{Scalar barrier and WKB transmission.}
		(a) Effective scalar potential for the Schwarzschild reference and the
		\(\epsilon=1100\) geometry.  (b) Corresponding WKB transmission factors
		versus scaled frequency.  The curves define the mean-branch filter; they
		are not exact greybody factors.}
\end{figure}

Supplemental Fig.~\ref{fig:S3-greybody}(a) shows that strong dressing changes
both the position and height of the scalar barrier.  The effect is modest for
\(\ell=0\) and becomes clearer for the centrifugal \(\ell=1,2\) barriers.
Panel (b) shows the corresponding consequence: relative to the Schwarzschild
curves, the strongly dressed transmission steps move to larger scaled
frequency.  Higher partial waves require larger \(\omega/T_{\eff}\) to become
transparent.  These changes determine the branch-dependent filter used in
the Letter.

\subsection{Filtered and unfiltered luminosity maps}

The unfiltered control shown in Fig.~\ref{fig:S2-correlation}(b) uses the
instantaneous Stefan--Boltzmann-type statistic
\begin{equation}
	L_{\eff,i}\propto A_{h,i}T_i^4
	=4\pi r_{h,i}^2T_i^4.
	\label{eq:unfiltered-luminosity}
\end{equation}
The proportionality constant cancels from a standardized moment.  This
control depends on both members of the stochastic pair and isolates the
nonlinear thermodynamic map before propagation through the exterior barrier.

For the greybody-filtered result in Fig.~3 of the Letter, the mean-branch
transmission is used:
\begin{equation}
	L_{\rm gb}(T_i)
	=\sum_{\ell=0}^{\ell_{\max}}\frac{2\ell+1}{2\pi}
	\int_0^\infty\dd\omega\,
	\Gamma_\ell^{\rm WKB}(\omega)
	\frac{\omega}{e^{\omega/T_i}-1}.
	\label{eq:luminosity}
\end{equation}
Within a fixed \(\epsilon\) branch,
\(\Gamma_\ell^{\rm WKB}\) is not fluctuated from sample to sample.
Stochasticity enters Eq.~\eqref{eq:luminosity} through \(T_i\).  The
calculation therefore does not include a stochastic scattering potential.

The origin of a positive luminosity skewness can be seen without assuming a
particular noise amplitude.  Let \(\theta=T-\bar T\) have zero mean and
variance \(\sigma_T^2\), and expand the filtered map locally:
\begin{equation}
	L_{\gb}(T)=L_{\gb}(\bar T)
	+a\theta+\frac{1}{2}b\theta^2+O(\theta^3),
	\quad
	a=L_{\gb}'(\bar T),\quad b=L_{\gb}''(\bar T).
	\label{eq:L-Taylor}
\end{equation}
If \(\theta\) is Gaussian at leading order, the centered luminosity
fluctuation is
\(\delta L_{\gb}=a\theta+\tfrac12b(\theta^2-\sigma_T^2)+\cdots\).  Its first
nonzero third central moment and standardized skewness are
\begin{align}
	\mu_3(L_{\gb})
	&=3a^2b\,\sigma_T^4+O(\sigma_T^6),
	\label{eq:L-third-moment}\\
	\gamma_1(L_{\gb})
	&=\frac{3b}{|a|}\sigma_T+O(\sigma_T^2).
	\label{eq:L-small-skew}
\end{align}
The Planck-weighted luminosity is increasing and locally convex over the
sampled temperature range, so \(a>0\) and \(b>0\); the leading skewness is
therefore positive.  Non-Gaussianity already generated by nonlinear root
tracking, and the additional dependence of the unfiltered control on
\(r_h\), supply further contributions beyond
Eq.~\eqref{eq:L-small-skew}.  The WKB barrier changes \(a\) and \(b\) through
its frequency weighting, but it is not required for the sign-generating
mechanism.

For either luminosity variable \(L_i\), define the central sample moments
\(m_k=N_{\rm valid}^{-1}\sum_i(L_i-\bar L)^k\) and
\(g_1=m_3/m_2^{3/2}\).  The adjusted Fisher--Pearson estimator is
\begin{equation}
	G_1=
	\frac{\sqrt{N_{\rm valid}(N_{\rm valid}-1)}}
	{N_{\rm valid}-2}\,g_1.
	\label{eq:skewness-corrected}
\end{equation}
This definition avoids mixing the \(1/N_{\rm valid}\) central moments used in
\(g_1\) with the \(1/(N_{\rm valid}-1)\) variance estimator used elsewhere.
Bootstrap error bars are obtained by resampling the retained luminosity
values and recalculating \(G_1\).  At \(\lambda_s=0\), \(\sigma_L=0\) and
skewness is undefined; the plotted zero is a continuous-limit convention.


\subsection{Interpretation of the luminosity curves}

Supplemental Fig.~\ref{fig:S2-correlation}(b) shows that the unfiltered
luminosity develops positive skewness on all branches once the noise is
finite.  This is expected from Eq.~\eqref{eq:unfiltered-luminosity}: the
fourth power of \(T_i\) weights upward temperature excursions much more
strongly than equal downward excursions.  The horizon factor
\(r_{h,i}^2\) competes with this thermal weighting, while the
horizon--temperature covariance determines whether it reinforces or reduces
the positive tail.  The \(\epsilon=700\) and \(1100\) curves reach their
enhanced regime earlier, consistent with their stronger joint response.

The greybody-filtered skewness in Fig.~3 of the Letter is smaller for the
weakly dressed branches and remains clearly positive for \(\epsilon=700\)
and \(1100\).  The difference between the filtered and unfiltered controls
does not mean that the WKB barrier creates the non-Gaussianity.  Rather, the
barrier changes the frequency weighting of the already nonlinear Planck map.
The dressed potentials in Fig.~\ref{fig:S3-greybody} suppress different
spectral regions, thereby changing both the magnitude and the branch ordering
of the standardized third moment.

The broad plateau at large \(\lambda_s\) combines the imposed nonlinear
amplitude map, the nonlinear temperature-to-luminosity transformation, and
conditional selection.  Once the temperature distribution is wide enough,
increasing its scale need not change a standardized shape parameter
proportionally.  At the same time, rejection of samples that fail the root or
temperature criteria limits the most extreme near-critical excursions.  A fully
time-dependent calculation with a fluctuating transmission potential could
modify this plateau, but that problem lies beyond the present effective
model.